\documentclass[aps,
prd,
nofootinbib,
showpacs,
twocolumn,
preprintnumbers,
floatfix]{revtex4-2}

\usepackage[bookmarks, linktocpage, colorlinks = true, linkcolor = blue, urlcolor  = blue, citecolor = blue, anchorcolor = green, hyperindex = true, hyperfigures]{hyperref}

\usepackage{graphicx} 
\usepackage{dcolumn}
\usepackage{multirow}
\usepackage{amsmath, amssymb}
\usepackage{slashed}
\usepackage[usenames]{color}
\usepackage{float}
\usepackage{bm}

\begin{document}

\preprint{RIKEN-iTHEMS-Report-26}

\title{Explaining the universal Hagedorn temperature in string models of hadrons with fixed endpoints or baryon junctions}

\author{Yuki Fujimoto}
\affiliation{Department of Physics, Niigata University, Ikarashi, Niigata 950-2181, Japan}
\affiliation{Center for Interdisciplinary Theoretical and Mathematical Sciences (iTHEMS), RIKEN, Wako 351-0198, Japan}

\date{\today}

\begin{abstract}
We derive the mass formulas and more accurate asymptotic expressions for the densities of states for string models of hadrons with one fixed heavy endpoint or a baryon junction.
We show that strings with one fixed endpoint have the same Hagedorn temperature as ordinary open strings at the same tension and number of polarizations, despite the different Regge slopes of heavy-light and light mesons.
This provides a microscopic explanation for the universality of the Hagedorn temperature recently found phenomenologically.
We also find that the baryon-junction and quark-diquark pictures share this Hagedorn temperature and yield the same asymptotic mass dependence of the density of states, differing only by an overall normalization.
The observed baryon mass spectrum provides little discrimination between the two pictures, although the baryon-junction model has a larger discrepancy with the observed Regge slopes.
\end{abstract}

\maketitle

\section{Introduction}

There has recently been renewed interest in string descriptions of hadronic spectra and hadron resonance gas thermodynamics~\cite{Fujimoto:2025sxx,Marczenko:2025nhj,Fujimoto:2025trl,Marczenko:2026yme,Marczenko:2026drt,Fujimoto:2026syp}, partly motivated by proposals for an intermediate regime of QCD matter between the hadronic phase and the quark-gluon plasma~\cite{Rohrhofer:2019qwq,Rohrhofer:2019qal,Glozman:2022lda,Cohen:2023hbq,Fujimoto:2025sxx,Hanada:2025rca}.
The connection between string excitations and an exponentially growing hadron spectrum has a long history, including dual-resonance models~\cite{Veneziano:1968yb,Fubini:1969qb,Huang:1970iq,Dienes:1993mp}.
The density of states grows exponentially with mass $m$ as $e^{m/T_H}$, where $T_H$ is the Hagedorn temperature.
While Hagedorn introduced such a spectrum through the statistical bootstrap~\cite{Hagedorn:1965st}, its string counterpart arises naturally from the many ways of exciting quantized transverse oscillations.
Phenomenological Hagedorn descriptions have often used $T_H \simeq 150$--$200\,\mathrm{MeV}$, close to the chiral crossover scale~\cite{Broniowski:2000bj,Broniowski:2004yh,Noronha-Hostler:2009wof,Majumder:2010ik,Cleymans:2011fx,Lo:2015cca,ManLo:2016pgd}.

In pure Yang-Mills theory, Meyer showed that a glueball gas supplemented by closed-string states reproduces the lattice entropy density in the confined phase, with a $T_H$ of order $300\,\mathrm{MeV}$, which is very close to the deconfinement temperature~\cite{Meyer:2009tq}.
Motivated by this result, we applied open-string spectra to hadrons with quark endpoints and found that a similar $T_H$ describes hadron resonance gas thermodynamics well~\cite{Fujimoto:2025sxx}.

Recent studies have described light-hadron spectra using this approach~\cite{Marczenko:2025nhj,Fujimoto:2025trl,Marczenko:2026yme} (see also Ref.~\cite{Dienes:1993mp} for earlier work) and extended it to heavy flavors~\cite{Marczenko:2026drt}.
Mesons are represented by quark-antiquark strings~\cite{Nambu:1974zg} and baryons by quark-diquark strings~\cite{Bars:1975dd,Selem:2006nd,Fujimoto:2025trl}.
After the quark-mass contributions were separated, a common $T_H \simeq 300\,\mathrm{MeV}$ was found to describe spectra across quark flavors as well as lattice-QCD thermodynamics in the charm sector~\cite{Marczenko:2026drt}.

The higher $T_H$ obtained in these string descriptions~\cite{Dienes:1993mp,Fujimoto:2025sxx,Marczenko:2025nhj} is consistent with the conventional scales $\mathcal{T}\sim1\,\mathrm{GeV}/\mathrm{fm}$ and $\alpha'\sim1\,\mathrm{GeV}^{-2}$~\cite{Lucini:2003zr,Selem:2006nd,Athenodorou:2020ani,Brambilla:2022het}, where $\mathcal{T}$ is the string tension and $\alpha'=1/(2\pi\mathcal{T})$ is the corresponding Regge slope.
For an ordinary open string in three spatial dimensions, string theory gives $T_H=\sqrt{\frac{3\mathcal{T}}{2\pi}} = \frac{1}{2\pi}\sqrt{\frac{3}{\alpha'}}$.

For heavy-light mesons, the effective Regge slope is twice the light-meson value in the heavy-quark limit~\cite{Olsson:1994dg,Selem:2006nd}, a behavior supported by phenomenology~\cite{Olsson:1994dg,Filipponi:1997vf,Sonnenschein:2014jwa,Chen:2017fcs}.
Naively inserting this slope into the ordinary open-string relation $T_H=\frac{1}{2\pi}\sqrt{\frac{3}{\alpha'}}$ would predict that $T_H$ for heavy-light mesons is $1/\sqrt{2}$ times that for light mesons.
This is at odds with the aforementioned common $T_H$ extracted phenomenologically~\cite{Marczenko:2026drt}.

A related question for baryons is whether replacing a quark-diquark string by three strings meeting at a junction changes the number and structure of the oscillator modes.
The quark-diquark picture is motivated by Regge phenomenology~\cite{Selem:2006nd}, while the junction picture has long been a subject of theoretical investigation~\cite{Artru:1974zn,Collins:1975yt,Kiriyama:1976ue,Sundermeyer:1976df,Rossi:1977cy,Asano:2022wou,Frenklakh:2023pwy,Altmann:2024odn,Komargodski:2024swh}.
Lattice-QCD calculations support a Y-shaped flux distribution at large quark separations~\cite{Takahashi:2000te,Takahashi:2002bw,Alexandrou:2002sn,Bissey:2006bz}.
Studies of baryon-number transport in nuclear collisions also support the junction picture~\cite{Kharzeev:1996sq,STAR:2024lvy}.
Comparing these two pictures tests whether the Hagedorn growth is sensitive to the geometry of the confining strings.

In this work, we derive the mass formulas and more accurate asymptotic expressions for the densities of states of quantized strings with one fixed heavy endpoint and with a baryon junction.
Counting the normal modes as independent bosonic oscillators, we find that both share the Hagedorn temperature of an ordinary open string at the same tension, despite their different Regge slopes.
This provides a microscopic explanation for the common $T_H$ found in Ref.~\cite{Marczenko:2026drt}.
We also show that the junction and quark-diquark spectra have the same asymptotic mass dependence and differ only by an overall normalization, making them difficult to distinguish using the available baryon mass spectrum.
We compare these predictions with classical Regge trajectories and cumulative light-meson, open-charm, and baryon spectra.
Using the more accurate asymptotic densities brings the Hagedorn temperatures extracted from different sectors into closer agreement.

The paper is organized as follows.
Section~\ref{sec:open} reviews the mass formula and the asymptotic density of states for an ordinary open string, and Secs.~\ref{sec:fixed} and \ref{sec:junction} treat fixed endpoints and baryon junctions.
Section~\ref{sec:hage} discusses Regge trajectories and the common Hagedorn temperature.
In Sec.~\ref{sec:fit}, we compare the models with cumulative hadron spectra and discuss the ideal-gas thermodynamics of the Bessel spectrum.

Throughout the paper, we label the spatial boundary conditions by NN and DN, with N and D denoting Neumann and Dirichlet conditions at free and fixed endpoints, respectively.
The letters refer to $\sigma=0$ and $\sigma=\pi$, in that order.
We adopt the mostly-plus convention for the spacetime metric.

\section{Mass spectrum of open strings}
\label{sec:open}

Here, we review the derivation of the mass formula and spectrum for an open NN string.
We follow the arguments found in the standard textbooks~\cite{Green:1987sp,Tong:2009np}.

\subsection{Mass formula and oscillator states}

A string sweeps out a worldsheet parametrized by a timelike coordinate $\tau$ and a spacelike coordinate $\sigma \in [0,\pi]$.
The embedding $X^\mu(\tau, \sigma)$ maps the worldsheet into $D$-dimensional Minkowski space, with $\mu = 0, \ldots, D-1$.
The Polyakov action is
\begin{align}
    S = - \frac1{4\pi \alpha'} \int d\tau d\sigma \sqrt{-g} g^{ab} \partial_a X^\mu \partial_b X_\mu\,,
\end{align}
where $g_{ab}$ is a dynamical metric on the worldsheet.
The stress-energy tensor is
\begin{align}
    T_{ab} = \partial_a X \cdot \partial_b X - \frac12 g_{ab} g^{cd} \partial_c X \cdot \partial_d X\,.
\end{align}
Varying the worldsheet metric imposes the constraints $T_{ab}=0$.
In conformal gauge, $g_{ab}=\eta_{ab}$, these become
\begin{align}
    T_{01} &= \dot X\cdot X'=0\,,\\
    T_{00}=T_{11} &= \frac12(\dot X^2+X'^2)=0\,.
    \label{eq:T00}
\end{align}
The equation of motion obtained by varying $X^\mu$ is the free wave
equation,
\begin{align}
    \ddot X^\mu-X''^\mu=0\,.
\end{align}

We impose the Neumann boundary conditions at both endpoints:
\begin{align}
    X'^\mu(\tau, 0) = X'^\mu(\tau, \pi) = 0\,.
\end{align}
Separating variables as $X(\tau, \sigma) = F(\tau) f(\sigma)$ reduces the wave equation to
\begin{align}
    \frac{F''(\tau)}{F(\tau)} = \frac{f''(\sigma)}{f(\sigma)} = -k^2\,.
\end{align}
For $k > 0$, the spatial part of the solution is
\begin{align}
    f(\sigma) = A \cos(k\sigma) + B \sin(k\sigma)\,.
\end{align}
From the boundary condition $f'(0) = 0$, it follows that $B=0$.
Further, from the boundary condition $f'(\pi) = 0$, it follows that $-Ak \sin(k\pi) = 0$.
Thus $k$ must be an integer: $k = n = 1, 2, \ldots$.
The spatial eigenmodes are $\cos(n \sigma)$.
The temporal part is
\begin{align}
    F_n(\tau) = A_n \cos(n\tau) + B_n \sin(n \tau)\,.
\end{align}
For the zero modes $f_0$ and $F_0$, the solutions satisfy
\begin{align}
    f_0'' = 0\,, \qquad
    F_0 '' = 0\,.
\end{align}
From the boundary condition, $f_0$ is a constant, and the temporal part becomes
\begin{align}
    F_0(\tau) = x + v \tau\,.
\end{align}
Then, the general solution is
\begin{align}
    X^\mu(\tau, \sigma) &= x^\mu + v^\mu \tau \\
    & \quad + \sum_{n=1}^\infty \left[A_n^\mu \cos(n\tau) + B_n^\mu \sin(n\tau) \right] \cos(n \sigma)\,. \notag
\end{align}
Note that the first two terms are $\sigma$-independent zero modes.

Let us now relate $v^\mu$ to the total momentum $p^\mu$.
The action after the gauge fixing in the flat metric is
\begin{align}
    S = \frac1{4\pi\alpha'} \int d\tau d\sigma (\dot{X}^2 - X'^2)\,.
\end{align}
The canonical momentum density is
\begin{align}
    \Pi_\mu = \frac{\partial \mathcal{L}}{\partial \dot{X}^\mu} = \frac{1}{2\pi\alpha'} \dot{X}_\mu\,.
\end{align}
The total momentum $p^\mu$ is the Noether charge associated with spacetime translations $X^\mu \to X^\mu + c^\mu$, obtained by integrating the canonical momentum density:
\begin{align}
    p^\mu = \int_0^\pi d\sigma \, \Pi^\mu = \frac1{2\pi\alpha'} \int_0^\pi d\sigma \, \dot{X}^\mu\,.
\end{align}
By substituting the general solution, the oscillatory part vanishes as the integral of $\cos(n\sigma)$ is zero, and we obtain
\begin{align}
    p^\mu = \frac{1}{2\pi\alpha'} \pi v^\mu\,.
\end{align}
It is convenient to combine $A_n^\mu$ and $B_n^\mu$ into complex coefficients,
\begin{align}
    \alpha_n^\mu = \frac{n}{2\sqrt{2\alpha'}}(B_n^\mu - iA_n^\mu)\,,\qquad
    \alpha_{-n}^\mu = (\alpha_n^\mu)^*\,.
\end{align}
Then the general solution becomes
\begin{align}
    X^\mu(\tau, \sigma) &= X_{\rm zm}^\mu + i \sqrt{2\alpha'} \sum_{n\neq 0} \frac{\alpha_n^\mu}{n} e^{-i n\tau}\cos(n \sigma)\,,
    \label{eq:modeexp}
\end{align}
where $X_{\rm zm}^\mu$ is the $\sigma$-independent zero mode
\begin{align}
    X_{\rm zm}^\mu \equiv x^\mu + 2 \alpha' p^\mu \tau\,.
\end{align}
The mass formula follows from the $\sigma$-integrated constraint \eqref{eq:T00}, which can be written in terms of the worldsheet Hamiltonian as
\begin{align}
    \mathcal H
    &= \int_0^\pi d\sigma\,
       \bigl(\Pi_\mu\dot X^\mu-\mathcal L\bigr) \notag\\
    &= \frac{1}{4\pi\alpha'}\int_0^\pi d\sigma\,
       \bigl(\dot X^2+X'^2\bigr)=0\,.
    \label{eq:Hopen}
\end{align}
Substituting the mode expansion \eqref{eq:modeexp} and using the orthogonality of the spatial mode functions gives
\begin{align}
    \mathcal{H} = \mathcal{H}_{\rm zm} + \mathcal{H}_{\rm osc}\,,
\end{align}
where $\mathcal{H}_{\rm zm}$ is the contribution of the $\sigma$-independent zero mode $X_{\rm zm}^\mu$, and $\mathcal{H}_{\rm osc}$ is the contribution of the nonzero oscillator modes.
The zero mode contribution is
\begin{align}
    \mathcal H_{\rm zm}=\frac{v^2}{4\alpha'}=\alpha'p^2\,,
\end{align}
while the nonzero modes give, classically,
\begin{align}
    \mathcal H_{\rm osc}
    =\sum_{n=1}^{\infty}\alpha_{-n}\cdot\alpha_n\,.
\end{align}

For quantization, we use the standard transverse formulation of the open string, with $b=D-2$ independent oscillator polarizations.
Their oscillators obey, for $n, m > 0$,
\begin{align}
    [\alpha_n^i,\alpha_{-m}^j]
    =n\delta_{nm}\delta^{ij}\,,\qquad i,j=1,\ldots,b\,.
\end{align}
Introducing creation and annihilation operators by
\begin{align}
    \alpha_n^i=\sqrt n\,a_{n,i}\,,\qquad
    \alpha_{-n}^i=\sqrt n\,a^\dagger_{n,i}\,,
\end{align}
each transverse mode contributes $n(a^\dagger_{n,i}a_{n,i}+1/2)$ to the oscillator Hamiltonian.
The excitation contribution has eigenvalue
\begin{align}
    N
    =\sum_{i=1}^{b}\sum_{n=1}^{\infty}n\,k_{n,i}\,,
    \label{eq:Npart}
\end{align}
where $k_{n,i}$ is the occupation number of mode $(n,i)$, i.e., $k_{n,i} |\psi \rangle = a^\dagger_{n,i}a_{n,i} |\psi \rangle$.
The value of $N$ is often called the level.
The vacuum energy is
\begin{align}
    \sum_{i=1}^{b}\sum_{n=1}^{\infty}\frac{n}{2} = \frac{b}{2} \zeta(-1)\,,
\end{align}
where we used the zeta-function regularization
\begin{align}
    \zeta(-1) = \sum_{n=1}^\infty n = -\frac1{12}\,.
\end{align}
Then, the integrated constraint on a physical state becomes
\begin{align}
    \mathcal H|\psi\rangle
    =\left(\alpha'p^2 + N - \frac{b}{24}\right)|\psi\rangle=0\,.
\end{align}
With $p^2=-m^2$, this yields the open-string mass formula
\begin{align}
    \alpha' m^2 = N - \frac{b}{24}\,.
    \label{eq:massNN}
\end{align}

\subsection{Integer partitions and the asymptotic density of states}

Let $p_b(N)$ denote the number of unordered partitions of the integer $N$ with $b$ independent polarizations.
Each partition is specified by non-negative occupation numbers $k_{n,i}$ satisfying Eq.~\eqref{eq:Npart}.
Summing independently over these occupation numbers gives the generating function
\begin{align}
    P_b(w) &\equiv \sum_{N=0}^\infty p_b(N) w^{N}
    = \prod_{n=1}^\infty (1 - w^n)^{-b}\,.
    \label{eq:Pbw}
\end{align}

We extract the large-$N$ coefficients using the small-$t$ asymptotic form.
Writing $w= e^{-t}\ (t >0)$ and expanding the logarithm gives
\begin{align}
    \log P_b(e^{-t}) = b \sum_{n=1}^\infty \sum_{k=1}^\infty \frac{e^{-nkt}}{k}\,.
\end{align}
To evaluate the double sum, we use the inverse Mellin transform
\begin{align}
    e^{-x} = \frac{1}{2\pi i} \int_{\gamma - i\infty}^{\gamma + i\infty}ds\, \Gamma(s) x^{-s}\,.
\end{align}
Substituting $x = nkt$ and choosing $\gamma > 1$ so that both sums converge, we obtain
\begin{align}
    \log P_b(e^{-t})
    = \frac{b}{2\pi i} \int ds \, \Gamma(s) t^{-s} \zeta(s) \zeta(s+1) \,.
\end{align}
The poles at $s = 1, 0, -1$ give the small-$t$ asymptotic form
\begin{align}
    \log P_b(e^{-t}) = \frac{b\pi^2}{6t} + \frac{b}2 \log \frac{t}{2\pi} - \frac{bt}{24} + O(t^2)\,.
    \label{eq:Pbasympt}
\end{align}
These terms arise from the poles at $s = 1, 0, -1$, respectively.
The same expansion follows from the modular transformation of the Dedekind eta function, using $P_b(e^{-t})=e^{-bt/24}\eta(it/2\pi)^{-b}$.
The resulting exact identity
\begin{align}
    P_b(e^{-t})
    &=
    \left(\frac{t}{2\pi}\right)^{b/2}
    \exp\left(\frac{b\pi^2}{6t}-\frac{bt}{24}\right)
    P_b(e^{-4\pi^2/t})
\end{align}
shows that the omitted terms in $\log P_b(e^{-t})$ are $O(e^{-4\pi^2/t})$ as $t \to 0$.

Applying Cauchy's integral formula to $P_b(w)$ and using the above asymptotic form gives
\begin{align}
    p_b(N) &\simeq \int \frac{dt}{2\pi i}\, \left(\frac{t}{2\pi}\right)^{b/2} \!\! \exp\left[\left(N - \tfrac{b}{24}\right)t + \tfrac{b\pi^2}{6t}\right]\,.
    \label{eq:pbN}
\end{align}
This is the integral representation of the modified Bessel function $I_{|\nu|}$ of order $|\nu|$, with $\nu \equiv -1 - b/2$.
Therefore, we obtain
\begin{align}
    p_b(N) &\simeq \frac{1}{(2\pi)^{b/2}}\left(\frac{N-\frac{b}{24}}{b\pi^2/6}\right)^{\nu/2}\notag \\
    & \qquad \times I_{|\nu|}\left(2\sqrt{\tfrac{b\pi^2}{6} \left(N - \tfrac{b}{24}\right)}\right)\,.
\end{align}

We now substitute the string mass formula \eqref{eq:massNN} into $p_b(N)$ obtained above to find the mass spectrum $\rho(m)$.
Namely,
\begin{align}
    \rho(m) dm \simeq p_b(N) dN\,.
\end{align}
Thus,
\begin{align}
    \rho(m) &\simeq
    \frac{2\alpha' m}{(2\pi)^{b/2}}\left(\frac{6\alpha' m^2}{b\pi^2}\right)^{\nu/2} I_{|\nu|}\left(2\pi\sqrt{\frac{b\alpha'}6}m\right)\,.
\end{align}
We define the Hagedorn temperature as
\begin{align}
    T_H = \frac{1}{2\pi} \sqrt{\frac{6}{b \alpha'}}\,,
    \label{eq:THopen}
\end{align}
and the asymptotic mass density of states of the bosonic open string is
\begin{align}
    \rho_{\rm open}(m)
    \simeq
    \frac{1}{T_H}
    \left(\frac{b\pi}{6}\right)^{\frac{b}2}
    \left(\frac{m}{T_H}\right)^{-\frac{b}2}
    I_{1+\frac{b}2}\left(\frac{m}{T_H}\right)\,.
    \label{eq:rhoopen}
\end{align}
Using the asymptotic expansion of $I_{\nu}(x)$ for large argument $x \gg 1$
\begin{align}
    I_\nu(x) \sim \frac{e^x}{\sqrt{2\pi x}} \left[1 - \frac{4\nu^2 - 1}{8x} + O(x^{-2})\right]\,,
    \label{eq:besselasymp}
\end{align}
we obtain
\begin{align}
    \rho_{\rm open}(m)
    &\sim
    \frac{1}{\sqrt{2\pi}T_H}
    \left(\frac{b\pi}{6}\right)^{\frac{b}2}
    \left(\frac{m}{T_H}\right)^{-\frac{b+1}2}
    e^{m/T_H} \notag \\
    &\quad \times \left[ 1 - \frac{(b+1)(b+3)}{8}\frac{T_H}{m} + O\left(\frac{T_H^2}{m^2}\right)\right]\,.
    \label{eq:rhoopenLO}
\end{align}
The leading term shows the Hagedorn-type exponential growth and is the asymptotic string density of states used in Refs.~\cite{Fujimoto:2025sxx,Marczenko:2025nhj,Fujimoto:2025trl,Marczenko:2026yme,Marczenko:2026drt,Fujimoto:2026syp}.


\section{Strings with a fixed heavy endpoint}
\label{sec:fixed}

We consider an open string with one static heavy endpoint and one massless free endpoint to model the heavy-light meson.
The worldsheet coordinates are $(\tau, \sigma)$ as above, with $\sigma \in [0,\pi]$.
As in the preceding sections, we adopt a simple free normal-mode counting prescription, retaining $b$ independent bosonic oscillator coordinates.
Note that the free oscillator counting used here and below does not constitute a full constrained quantization; see, e.g., Refs.~\cite{Gursoy:2001bb,Asano:2022wou} for fixed-endpoint strings and junctions, respectively.

\subsection{Boundary conditions and the energy formula}

The heavy endpoint's rest-energy contribution follows from its point-particle action,
\begin{align}
    S_Q = - m_Q \int d \ell_Q\,,
    \qquad d \ell_Q = \sqrt{- dX_Q^\mu dX_{Q\mu}}\,.
\end{align}
In target-space time $t=X^0$, its Lagrangian and energy are
\begin{align}
    L_Q &= - m_Q \sqrt{1 - \boldsymbol{v}_Q^2}\,, \\
    E_Q &= \boldsymbol{p}_Q \cdot \boldsymbol{v}_Q - L_Q
    = \frac{m_Q}{\sqrt{1 - \boldsymbol{v}_Q^2}}\,.
\end{align}
For a static endpoint, $E_Q = m_Q$.
Additivity of the endpoint and string contributions to the conserved energy therefore gives
\begin{align}
    E = m_Q + \varepsilon\,,
\end{align}
where $\varepsilon$ is the string energy measured in the rest frame of the static endpoint.

With the heavy endpoint at the spatial origin, we impose the following spatial boundary conditions:
\begin{align}
    X^j(\tau, 0) = X'^j(\tau, \pi) = 0\,,
    \qquad
    j = 1, \ldots, D-1\,.
\end{align}
The spatial coordinates obey Dirichlet and Neumann boundary conditions at $\sigma = 0$ and $\pi$, respectively.
Solving the wave equation with these boundary conditions gives the spatial mode functions
\begin{align}
    f_r(\sigma)\propto\sin(r\sigma)\,,\qquad
    r=\frac12,\frac32,\frac52,\ldots\,.
\end{align}
The half-integer values of $r$ follow from the free-end condition $f_r'(\pi)=0$, which requires $\cos(r\pi)=0$.
The fixed endpoint eliminates the spatial translational zero mode.

The temporal coordinate is not fixed at the heavy endpoint, whose worldline extends along the time direction.
It obeys Neumann boundary conditions at both endpoints,
\begin{align}
    X'^0(\tau, 0) = X'^0(\tau, \pi) = 0\,,
\end{align}
and therefore admits a $\sigma$-independent zero mode.
In conformal gauge, we choose the temporal gauge $X^0=t_0+v^0\tau$.
The canonical momentum density then gives the string energy:
\begin{align}
    \varepsilon = -\int_0^\pi d\sigma\,\Pi_0
    = \frac{v^0}{2\alpha'}\,.
    \label{eq:e_v0}
\end{align}
Using the oscillator normalization derived above, we define the level as
\begin{align}
    N_{\rm DN} = \sum_{i=1}^b\sum_{r=\frac12,\frac32,\ldots}r\,k_{r,i}\,,
\end{align}
where the subscript $\mathrm{DN}$ denotes the string with Dirichlet and Neumann conditions at fixed and free endpoints, respectively.
Since $N_{\rm DN}$ can take half-integer values, it is convenient to introduce the integer level
\begin{align}
    K_{\rm DN} = 2 N_{\rm DN}\,.
\end{align}
The vacuum energy contribution is
\begin{align}
    \sum_{i=1}^b\sum_{r=\frac12,\frac32,\ldots}\frac{r}{2} = \frac{b}2 \zeta\left(-1,\frac12 \right)\,,
\end{align}
where we used the zeta-function regularization
\begin{align}
    \zeta\left(-1,\frac12 \right) = \sum_{k = 0}^\infty \left(k + \frac12\right) = \frac1{24}\,.
\end{align}

To relate the excitation level to the string energy, we use the vanishing worldsheet Hamiltonian condition \eqref{eq:Hopen}.
This integration selects its Fourier zero component.
Although the spatial coordinates have no translational zero modes, their quadratic oscillator terms contribute to this component, yielding $N_{\rm DN}$ together with a vacuum contribution $b/48$.
The integrated constraint therefore reads
\begin{align}
    0 &= \frac{1}{4\pi\alpha'} \int_0^\pi d\sigma\, \left(\dot{X}^2 + X'^2 \right) \notag \\
    &=-\alpha' \varepsilon^2 + N_{\rm DN} + \frac{b}{48}\,,
\end{align}
where the first term in the last line arises from the integral of $\dot X^0\dot X_0 = -(v^0)^2$.
The energy formula for the integer level $K_{\rm DN}$ is
\begin{align}
    \alpha' \varepsilon^2 = \frac{K_{\rm DN}}2 + \frac{b}{48}\,.
    \label{eq:eneDN}
\end{align}
The relation of this energy formula to the classical Regge trajectory will be discussed in Sec.~\ref{sec:Regge}.
Here, $K_{\rm DN}$ labels the excitation level rather than the angular momentum $J$.
Also note that the sign of the vacuum energy contribution is opposite to the NN open string~\eqref{eq:massNN}.

\subsection{State counting and Hagedorn growth} 

We now count the oscillator states.
We define $g_{\rm DN}(K_{\rm DN})$ as the number of string oscillator states at the integer level $K_{\rm DN}$, with its dependence on $b$ left implicit.
Since each oscillator contributes an odd integer $2r$ to $K_{\rm DN}$, its generating function is
\begin{align}
    G_{\rm DN}(w)
    &\equiv \sum_{K_{\rm DN}=0}^\infty g_{\rm DN}(K_{\rm DN}) w^{K_{\rm DN}}\\
    &=\prod_{n=1}^\infty(1 - w^{2n-1})^{-b}\,.
\end{align}
The mode weights $2r$ are odd, but all nonnegative integer levels are allowed for $K_{\rm DN}$.
Using $P_b(w)$ defined in Eq.~\eqref{eq:Pbw}, $G_{\rm DN}(w)$ can be expressed as
\begin{align}
    G_{\rm DN}(w) 
    =\frac{P_b(w)}{P_b(w^2)}\,.
\end{align}
Setting $w = e^{-t}$ and using the expansion of $P_b$~\eqref{eq:Pbasympt} obtained above yields
\begin{align}
    G_{\rm DN}(e^{-t})
    \simeq 2^{-b/2}\exp\left(\frac{b\pi^2}{12t} + \frac{bt}{24}\right)\,
\end{align}
where the powers of $t$ cancel in the ratio.
Applying the same integral representation for the modified Bessel function used above for $p_b(N)$, we obtain
\begin{align}
    g_{\rm DN}(K_{\rm DN}) &\simeq
    \frac{1}{2^{b/2}}\left(\frac{b\pi^2/12}{K_{\rm DN}+\frac{b}{24}}\right)^{1/2}\notag \\
    & \qquad \times I_{1}\left(2\sqrt{\tfrac{b\pi^2}{12} \left(K_{\rm DN} + \tfrac{b}{24}\right)}\right)\,.
\end{align}

We now convert $g_{\rm DN}(K_{\rm DN})$ to a smoothed density per unit string energy $\rho_{\rm DN}(\varepsilon)$.
Substituting the energy formula at large level \eqref{eq:eneDN} into the formula above,
we obtain the energy spectrum including the Jacobian prefactor
\begin{align}
    \rho_{\rm DN}(\varepsilon) \simeq
    \frac{1}{2^{b/2} T_H} I_1 \left(\frac{\varepsilon}{T_H}\right)\,,
    \label{eq:rhofixed}
\end{align}
where the Hagedorn temperature is defined as
\begin{align}
    T_H=\frac1{2\pi}\sqrt{\frac6{b\alpha'}}\,.
    \label{eq:THfixed}
\end{align}
For the same $\alpha'$ and $b$, the Hagedorn temperature agrees with that of the ordinary open string~\eqref{eq:THopen}.

Applying the asymptotic expansion \eqref{eq:besselasymp} to the equation above yields
\begin{align}
    \rho_{\rm DN}(\varepsilon) &\sim
    \frac{2^{-b/2}}{\sqrt{2\pi}\,T_H}
    \left(\frac{\varepsilon}{T_H}\right)^{-1/2}
    e^{\varepsilon/T_H}\notag\\
    &\quad \times \left[1 - \frac38 \frac{T_H}{\varepsilon} + O\left(\frac{T_H^2}{\varepsilon^2}\right) \right]\,.
    \label{eq:rhofixedLO}
\end{align}
The power of the prefactor is instead $-1/2$, compared with $-(b+1)/2$ for an open string.


\section{Strings with baryon junctions}
\label{sec:junction}

We consider three string arms with equal string tension $1/(2\pi\alpha')$ joined at a massless junction, with free outer endpoints.
For state counting, we treat the normal modes as noninteracting bosonic oscillators with $b$ independent polarizations as above.

\subsection{Junction conditions and the mass formula}

The derivation has two ingredients: the junction boundary conditions determine the oscillator frequencies, while the normalization of the total target-space momentum determines the mass formula.

Let $X_A^\mu(\tau, \sigma)$, $A=1,2,3$, describe the arms, with $0\le \sigma \le\pi$.
In conformal gauge, the action is
\begin{align}
    S=\frac{1}{4\pi\alpha'}
    \sum_{A=1}^{3}\int d\tau\int_0^\pi d \sigma \,
    \left(\dot X_A^2-X_A'^2\right)\,.    
\end{align}
The junction is at $\sigma = 0$, where
\begin{align}
    X_1(\tau,0)=X_2(\tau,0)=X_3(\tau,0)\,,\\
    X_1'(\tau,0)+X_2'(\tau,0)+X_3'(\tau,0)=0\,.   
\end{align}
At the free outer endpoints,
\begin{align}
    X_A'(\tau,\pi)=0.
\end{align}
These are the classical continuity and force-balance conditions for a junction with equal tension.
We introduce the orthonormal combinations
\begin{align}
    U &= \frac{X_1 + X_2 + X_3}{\sqrt3}\,,\\
    V &= \frac{X_1 - X_2}{\sqrt2}\,,\\
    W &= \frac{X_1 + X_2 - 2X_3}{\sqrt6}\,.
\end{align}
Because the transformation is orthogonal, the quadratic action becomes a sum of three separate actions for $U,V,W$.
For the symmetric coordinate $U$, force balance gives
\begin{align}
    U'(\tau,0) = U'(\tau,\pi)=0\,.
\end{align}
Thus its spatial eigenfunctions are
\begin{align}
    \cos(n\sigma)\,,\qquad n=0,1,2,\ldots\,.
\end{align}
For the relative coordinates, continuity gives
\begin{align}
    V(\tau,0)=W(\tau,0)=0\,,
\end{align}
whereas the outer endpoints satisfy
\begin{align}
    V'(\tau,\pi)=W'(\tau,\pi)=0\,.
    \label{eq:VWfreeend}
\end{align}
Their eigenfunctions are
\begin{align}
    \sin(r \sigma)\,, \qquad r = \frac12, \frac32, \frac52, \ldots\,.
\end{align}
The half-integer values of $r$ follow from the free-end condition \eqref{eq:VWfreeend}, which requires $\cos(r\pi)=0$.

Consequently, the connected junction has one integer-frequency tower and two half-integer-frequency towers.
These are independent normal modes of the connected system.
To obtain the mass formula, first consider the common translational zero mode,
\begin{align}
    X_{A,\mathrm{zm}}^\mu=x^\mu+v^\mu\tau\,.
\end{align}
The canonical momentum density is
\begin{align}
    \Pi_A^\mu=\frac{\dot X_A^\mu}{2\pi\alpha'}\,,
\end{align}
so the total conserved momentum is
\begin{align}
    P^\mu=\sum_{A=1}^{3}\int_0^\pi d\sigma \,\Pi_A^\mu
    =\frac{3v^\mu}{2\alpha'}\,.
\end{align}
Only the symmetric zero mode contributes: the relative modes cancel in the sum over arms, and the nonzero symmetric modes integrate to zero.
Therefore,
\begin{align}
    v^\mu=\frac{2\alpha'}{3}P^\mu\,.
\end{align}
The integrated world-sheet constraint is
\begin{align}
    \mathcal{H} =
    \frac{1}{4\pi\alpha'}
    \sum_{A=1}^{3}\int_0^\pi d\sigma \,
    \left(\dot X_A^2+X_A'^2\right)=0\,.
\end{align}
Its translational contribution is
\begin{align}
    \mathcal{H}_{\rm zm}
    =\frac{3}{4\alpha'}v^2
    =\frac{\alpha'}{3}P^2.
\end{align}
Using the mostly-plus convention, $P^2=-m^2$, this becomes
\begin{align}
    \mathcal{H}_{\rm zm}=-\frac{\alpha'm^2}{3}\,.
\end{align}
Each nonzero normal mode contributes its frequency times its occupation number.
We define the excitation level by
\begin{align}
    N_{Y}=
    \sum_{i=1}^{b}
    \left(
    \sum_{n=1}^{\infty}n\,k_{n,i}
    +
    \sum_{\lambda=1}^{2}
    \sum_{r=\frac12,\frac32,\ldots}
    r\,\ell_{r,\lambda,i}
    \right)\,.
\end{align}
Here $\lambda$ labels the two relative-mode towers and the subscript $Y$ denotes the string with a Y-shaped baryon junction.
The vacuum energy is
\begin{align}
    &\sum_{i=1}^{b}
    \left(
    \sum_{n=1}^{\infty}\frac{n}2
    +
    \sum_{\lambda=1}^{2}
    \sum_{r=\frac12,\frac32,\ldots}
    \frac{r}2
    \right) \notag\\
    &= \frac{b}2 \left[\zeta(-1) + 2 \zeta\left(-1,\frac12\right)\right] = 0 \,,
\end{align}
where we used the zeta-function regularization.

The constraint $\mathcal{H}_{\rm zm} + \mathcal{H}_{\rm osc} = 0$ becomes
\begin{align}
    -\frac{\alpha'm^2}{3} + N_{Y} = 0\,.
\end{align}
Hence, in this model,
\begin{align}
    \alpha'm^2=3N_{Y}\,.
    \label{eq:massY}
\end{align}
Since $N_{Y}$ can take half-integer values, it is convenient to introduce the integer level
\begin{align}
    K_{Y} = 2 N_{Y}\,.
\end{align}
For the state counting below, we will use $K_{Y}$ instead of $N_{Y}$.

\subsection{State counting and the Hagedorn growth}

We now count the oscillator states.
We define $g_{Y}(K_{Y})$ as the number of transverse oscillator states at the integer level $K_{Y}$, with its dependence on $b$ left implicit.
Because an integer-frequency oscillator contributes $2n$ to $K_{Y}$ and a half-integer-frequency oscillator contributes an odd integer $2n-1$, its generating function is
\begin{align}
    G_{Y}(w)
    &\equiv \sum_{K_{Y}=0}^{\infty} g_{Y}(K_{Y}) w^{K_{Y}} \\
    &=\prod_{n=1}^{\infty} (1 - w^{2n})^{-b}
    \prod_{n=1}^{\infty} (1 - w^{2n-1})^{-2b}\,.
\end{align}
With $P_b(w)$ defined in Eq.~\eqref{eq:Pbw}, this can be rewritten as
\begin{align}
    G_{Y}(w) = P_b(w^2) \left(\frac{P_b(w)}{P_b(w^2)}\right)^2 = \frac{P_b(w)^{2}}{P_b(w^2)}\,.
\end{align}
Now we set $w = e^{-t}$.
Using the asymptotic expansion of $P_b$~\eqref{eq:Pbasympt}, we find
\begin{align}
    G_{Y}(e^{-t})
    \sim
    \left(\frac{t}{4\pi}\right)^{b/2}
    \exp\left(\frac{b\pi^2}{4t}\right)\,.
\end{align}
Applying the same integral representation for the modified Bessel function used above for $p_b(N)$ and $g_{\rm DN}(K_{\rm DN})$, we obtain
\begin{align}
    g_{Y}(K_{Y})
    &\simeq \frac{1}{(4\pi)^{b/2}}\left(\frac{K_Y}{\tfrac{b\pi^2}{4}}\right)^{\nu/2} I_{|\nu|}\left(2\sqrt{\tfrac{b\pi^2}{4} K_Y}\right)\,.
\end{align}

Finally, using the mass formula,
\begin{align}
    K_{Y} = \frac{2\alpha'm^2}{3}\,,
\end{align}
we convert the level degeneracy into a smoothed density per unit mass via the relation
\begin{align}
    \rho_{Y}(m)\,dm\simeq g_{Y}(K_{Y})\,dK_{Y}\,.
\end{align}
We obtain
\begin{align}
    \rho_Y(m) \simeq
    \frac{1}{T_H}
    \left(\frac{b\pi}{8}\right)^{\frac{b}2}
    \left(\frac{m}{T_H}\right)^{-\frac{b}2}
    I_{1+\frac{b}2}\left(\frac{m}{T_H}\right)\,,
    \label{eq:rhoY}
\end{align}
where the Hagedorn temperature is defined as
\begin{align}
    T_H = \frac{1}{2\pi}\sqrt{\frac{6}{b\alpha'}}\,.
\end{align}
For the same $\alpha'$ and $b$, the Hagedorn temperature agrees with that of the ordinary open string~\eqref{eq:THopen} and the string with one fixed endpoint~\eqref{eq:THfixed}.

Applying the asymptotic expansion \eqref{eq:besselasymp} to the equation above yields
\begin{align}
    \rho_{Y}(m) &\simeq
    \frac{1}{\sqrt{2\pi}\,T_H}
    \left(\frac{b\pi}{8}\right)^{\frac{b}2}
    \left(\frac{m}{T_H}\right)^{-\frac{b+1}2}
    e^{m/T_H}\notag\\
    &\quad \times \left[ 1 - \frac{(b+1)(b+3)}{8}\frac{T_H}{m} + O\left(\frac{T_H^2}{m^2}\right)\right]\,.
    \label{eq:rhoYLO}
\end{align}
By comparing the asymptotic density of states of baryon junctions with that of open strings \eqref{eq:rhoopen}, we find
\begin{align}
    \rho_{Y}(m) = \left(\frac34\right)^{b/2}{\rho_{\rm open}(m)}\,.
\end{align}
The Hagedorn temperature and the power of $m$ agree.
The additional oscillator towers change the growth with excitation level, but the factor of three in the mass formula \eqref{eq:massY} compensates for that change in the leading mass-dependent exponential.


\section{Regge trajectories and Hagedorn temperatures}
\label{sec:hage}

The classical rotating solutions relate energy to angular momentum, whereas the density of states counts all oscillator excitations at a given energy.
We first relate the angular momentum of a rotating string to its excitation level, and then explain why NN, DN, and junction strings share the same Hagedorn temperature despite their different Regge slopes.

\subsection{Classical Regge trajectories and quantum mass formulas}
\label{sec:Regge}

The classical rotating-string picture provides a natural starting point for comparing the ordinary open NN string with the DN string having one fixed heavy endpoint.
For the same string tension $\mathcal{T} = 1/(2\pi\alpha')$, their classical leading trajectories are~\cite{Olsson:1994dg,Selem:2006nd}
\begin{align}
    E_{\rm NN}^2 &= \frac{J_{\rm NN}}{\alpha'}\,,\\
    (E_{\rm DN} - m_Q)^2 &= \frac{J_{\rm DN}}{2\alpha'}\,.
\end{align}
The first NN relation assumes two massless endpoints, whereas the second DN relation applies in the static heavy-endpoint limit with the other endpoint massless.
We will recover these relations from the rotating solutions below and clarify their correspondence with the quantum excitation levels.
The effective Regge slope is therefore $2\alpha'$ for DN, twice the NN value.

The quantum mass formulas instead relate energy to the excitation level, which is a frequency-weighted sum of oscillator occupation numbers.
The relation between this level and angular momentum must therefore be established using the rotating solutions.

This relation between $N_{\rm DN}$ and $J_{\rm DN}$ can be obtained directly from the classical rotating solution.
The classical quantity corresponding to $N_{\rm DN}$ is the spatial oscillator Hamiltonian.
For motion in the $X^1\!-\!X^2$ plane, it is
\begin{align}
    N_{\rm DN}^{\rm cl} = \frac{1}{4\pi\alpha'}
    \int_0^\pi d\sigma\, \sum_{i=1}^{2}
    \left[(\dot X^i)^2 + (X'^i)^2\right]\,.
\end{align}
Quantizing this Hamiltonian gives the oscillator contributions $r(k_{r,i}+1/2)$, and removing the vacuum contribution yields the level $N_{\rm DN}=\sum_{r,i}r k_{r,i}$ used above.
The angular momentum about the fixed endpoint is instead the spatial rotation charge,
\begin{align}
    J &= \int_0^\pi d\sigma\,
      \left(X^1 \Pi_2 - X^2 \Pi_1\right) \notag\\
    &=\frac{1}{2\pi\alpha'} \int_0^\pi d\sigma\,
      \left(X^1 \dot{X}^2 - X^2 \dot{X}^1\right)\,.
\end{align}
These definitions apply to both NN and DN strings; the boundary conditions determine the mode functions substituted into them.

To construct a rotating solution, we take $X^0 = v^0 \tau$ and a single spatial eigenmode,
\begin{align}
    \boldsymbol X(\tau,\sigma)
    =f(\sigma)\left(\boldsymbol a\cos\omega\tau
                     +\boldsymbol b\sin\omega\tau\right)\,,
\end{align}
where $f'' + \omega^2f=0$.
The constraint $\dot X\cdot X'=0$ requires the length of the vector in parentheses to be constant, giving $|\boldsymbol{a}| = |\boldsymbol{b}|$ and $\boldsymbol{a} \cdot \boldsymbol{b} = 0$.
Choosing coordinate axes along these vectors gives
\begin{align}
    (X^1, X^2)=Af(\sigma)
       (\cos\omega\tau, \sin\omega\tau)\,.
\end{align}
The remaining constraint $\dot X^2+X'^2=0$ fixes $v^0=A\omega$ for $f(\sigma)=\cos\omega\sigma$ or $\sin\omega\sigma$.
This describes a straight string rotating uniformly in the $X^1\!-\!X^2$ plane.

For the DN string, the lowest mode with $\omega = 1/2$ gives the rigidly rotating solution
\begin{align}
    X^0 &= \frac A2\tau\,,\\
    X^1 &= A\sin\frac{\sigma}{2}\cos\frac{\tau}{2}\,,\\
    X^2 &= A\sin\frac{\sigma}{2}\sin\frac{\tau}{2}\,.
\end{align}
Substituting these equations into Eq.~\eqref{eq:e_v0} and the relations above, we obtain
\begin{align}
    \varepsilon &= \frac{A}{4\alpha'}\,, &
    N_{\rm DN}^{\rm cl} &= \frac{A^2}{16\alpha'}\,, &
    J &= \frac{A^2}{8\alpha'}\,.
\end{align}
Thus, for this rotating configuration,
\begin{align}
    \alpha'\varepsilon^2 = N_{\rm DN}^{\rm cl} = \frac{J}2\,.
\end{align}
The energy formula expressed through the excitation level and the classical Regge formula are therefore consistent:
the factor of $1/2$ arises from the relation between excitation level and angular momentum, already at the classical level.

For comparison, the lowest rotating NN mode with $\omega = 1$ is
\begin{align}
    X^0&=A\tau\,,\\
    X^1&=A\cos\sigma\cos\tau\,,\\
    X^2&=A\cos\sigma\sin\tau\,.
\end{align}
Using the same definitions gives
$N^{\rm cl} = J = A^2 / (4\alpha')$ and $E_{\rm NN} = A / (2\alpha')$, hence $\alpha'E_{\rm NN}^2 = N^{\rm cl} = J$.
For a rotating mode of frequency $\omega$, the definitions above give $N^{\rm cl}=\omega J$.
The fundamental frequencies are $1$ for NN and $1/2$ for DN, explaining the different relations between excitation level and angular momentum.
Nevertheless, the energy formulas expressed in terms of level have the same form at large level:
\begin{align}
    \alpha'E_{\rm NN}^2\simeq N\,,\qquad
    \alpha'\varepsilon^2\simeq N_{\rm DN}\,.
\end{align}

In oscillator language, each quantum with a given circular polarization carries one unit of angular momentum and contributes $\omega$ to the level.
At fixed level, the largest angular momentum is therefore obtained by placing all quanta in the lowest-frequency mode with the same circular polarization.
This gives $J_{\max}=N$ for NN and $J_{\max}=2N_{\rm DN}$ for DN.
These maximum-spin states form the leading Regge trajectories.

The rotating Y junction is obtained by rotating the DN solution in the $X^1\!-\!X^2$ plane:
\begin{align}
    X_a^0 &= \frac A2\tau\,,\\
    X_a^1+iX_a^2 &= A\sin\frac{\sigma}{2}\,e^{i(\tau/2+\phi_a)}\,,
\end{align}
where $\phi_a=2\pi(a-1)/3$, $a=1,2,3$.
Its symmetric spatial coordinate vanishes, while the relative coordinates are proportional to $\sin(\sigma/2)$ and satisfy the DN conditions.
Summing the energies, spatial oscillator Hamiltonians, and angular momenta of the three arms gives
\begin{align}
    m_Y &= \frac{3A}{4\alpha'}\,, &
    N_Y^{\rm cl} &= \frac{3A^2}{16\alpha'}\,, &
    J_Y &= \frac{3A^2}{8\alpha'}\,.
\end{align}
Hence $J_Y=2N_Y^{\rm cl}$ and $\alpha'm_Y^2=3N_Y^{\rm cl}=3J_Y/2$, consistent with Eq.~\eqref{eq:massY} and giving the classical Regge slope $2\alpha'/3$.

\subsection{Universality of the Hagedorn temperature}

The NN and DN strings have the same energy formula at large level.
Their Hagedorn temperatures therefore agree if the state counts have the same leading exponential growth as functions of excitation level.
For the same $\alpha'$ and $b$, this follows from the spacing of their oscillator frequencies:
\begin{align}
    \omega_{\rm NN}&=1,2,3,\ldots\,,\\
    \omega_{\rm DN}&=\frac12,\frac32,\frac52,\ldots\,.
\end{align}
Both frequency sequences have unit spacing, so they have the same asymptotic frequency density.
We briefly revisit the derivation of the density of states to make this connection explicit.

The total level is the sum of these frequencies weighted by their occupation numbers.
Thus $N$ is integer-valued, while $N_{\rm DN}$ can be integer or half-integer.
The integer label $K_{\rm DN}=2N_{\rm DN}$ used above rescales the level, but it does not change the physical frequency spacing.

To connect this interpretation with the preceding derivations, let $Z_{\rm NN}(t)$ and $Z_{\rm DN}(t)$ count states with weights $e^{-tN}$ and $e^{-tN_{\rm DN}}$, respectively.
We note that they connect to the generating functions used above as $Z_{\rm NN}(t)=P_b(e^{-t})$ and $Z_{\rm DN}(t) = G_{\rm DN}(e^{-t/2})$.
The argument $e^{-t/2}$ accounts for $K_{\rm DN}=2N_{\rm DN}$.
Likewise, for the Y junction, $Z_{Y}(t) = G_{Y}(e^{-t/2})$, since $K_{Y} = 2N_{Y}$.
These are generating functions for excitation levels.
For either frequency sequence, the logarithm is a sum over modes,
\begin{align}
    \log Z(t) = -b\sum_\omega \log(1 - e^{-t\omega})\,.
\end{align}
At large excitation level, the relevant limit is $t\to0$.
Replacing the sum by an integral, the unit frequency spacing yields the same measure $d\omega$ in both cases, giving
\begin{align}
    \log Z_{\rm NN}(t)
    \sim \log Z_{\rm DN}(t)
    &\sim -b \int_0^\infty d\omega\,
        \log(1 - e^{- t\omega}) \notag\\
    &=\frac{c}{t}\,,
\end{align}
where $c = b\pi^2 / 6$.
The saddle-point exponent for the NN string in Eq.~\eqref{eq:pbN} is $Nt + c / t$, giving the leading exponential growth $e^{2\sqrt{cN}}$ of the state count in the saddle-point approximation.
The same holds with $N$ replaced by $N_{\rm DN}$.
Using the common energy formula $\alpha' m^2 \simeq N$ or $\alpha' \varepsilon^2 \simeq N_{\rm DN}$ therefore gives
\begin{align}
    \log\rho_{\rm NN}(m)&\sim2\sqrt{c N}
        =2\sqrt{c \alpha'}\,m\,,\\
    \log\rho_{\rm DN}(\varepsilon)&\sim2\sqrt{c N_{\rm DN}}
        =2\sqrt{c \alpha'}\,\varepsilon\,.
\end{align}
The two spectra thus have the same Hagedorn temperature.
The frequency shift changes subleading terms and hence the algebraic prefactors in Eqs.~\eqref{eq:rhoopen} and \eqref{eq:rhofixed}, while leaving this leading exponential growth unchanged.

For a junction, increasing the number of arms increases the number of oscillator modes, but also changes the relation between total excitation level and physical mass.
Consider $\nu$ equal-tension arms, each parametrized by $0\leq\sigma\leq\pi$, joined at a massless junction with free outer endpoints.
The junction conditions give one integer-frequency tower and $\nu-1$ half-integer-frequency towers.
Since every tower has unit frequency spacing, the combined mode density is $\nu$ times that of a single string.
The generating function weighted by the total excitation level $N_\nu$ therefore has the leading behavior $\log Z_\nu(t)\sim\nu c/t$.
For $\nu=3$, this is precisely the leading term of $Z_{Y}(t)=G_{Y}(e^{-t/2})$.

The normalization of the common translational zero mode also involves the sum over all arms.
Generalizing the mass relation \eqref{eq:massY} and using the leading exponential growth obtained from the saddle-point counting gives
\begin{align}
    \alpha'm^2 &\simeq \nu N_\nu\,,\\
    \log\rho_\nu(m)
    &\sim 2\sqrt{\nu c N_\nu}
    =2\sqrt{c\alpha'}\,m\,.
\end{align}
Thus, the additional modes increase the density at fixed excitation level, while the same sum over arms changes the excitation level corresponding to a fixed physical mass.
The number of arms cancels in the leading exponential growth per unit mass, yielding
\begin{align}
    T_H=\frac{1}{2\sqrt{c\alpha'}}
    =\frac{1}{2\pi}\sqrt{\frac{6}{b\alpha'}}\,.
\end{align}

We note that closed strings used to model glueballs also share this Hagedorn temperature at the same string tension and polarization count~\cite{Meyer:2009tq,Fujimoto:2025sxx}.

The rotating trajectories depend on the frequencies of the modes used to carry angular momentum, while the leading growth of the total state count depends on the asymptotic density of oscillator frequencies.
Boundary conditions can therefore change the Regge slopes and spectral prefactors while leaving the Hagedorn temperature unchanged.

\section{Phenomenological applications}
\label{sec:fit}

We now compare the string spectra with cumulative hadron spectra from the Particle Data Group (PDG) 2024 tables~\cite{ParticleDataGroup:2024cfk}.
We first examine how retaining the Bessel form of the density changes the Hagedorn temperature extracted from light mesons and baryons.
We then compare the fixed-endpoint and ordinary open-string descriptions of open charm, focusing on their consistency with the temperature scale found in the light sectors.
We next compare the baryon-junction and quark-diquark descriptions and their implications for Regge trajectories.
Finally, we give an analytic illustration of the ideal-gas thermodynamics of the Bessel spectrum.
Throughout this section, we take $b=2$ and retain the absolute normalization of each string density.
The thresholds and total channel multiplicities used in all fits are summarized in Table~\ref{tab:stringchannels}.
The Bessel densities are used in the fits, with their leading large-mass terms retained for comparison in Sec.~\ref{sec:besselfit}.

We treat each resonance as a particle state neglecting its width, and construct the PDG cumulative spectrum by summing its spin degeneracy $2J+1$ over the included charge states.
The model parameters are obtained by minimizing the root mean square error (RMSE) of the logarithm of the cumulative count,
\begin{align}
    \mathrm{RMSE}_{\ln N}
    = \sqrt{\frac{1}{n_{\rm fit}} \sum_{k = 1}^{n_{\rm fit}} \left(\ln\frac{N_k^{\rm model}}{N_k^{\rm PDG}}\right)^2}\,.
    \label{eq:fitrmse}
\end{align}
Here, $n_{\rm fit}$ is the number of selected fit points, and $N_k^{\rm model}$ and $N_k^{\rm PDG}$ are the model and PDG cumulative counts at the same point.
Each pair of models is compared using the same data points and equal weights in Eq.~\eqref{eq:fitrmse}.
The cumulative counts retain all states below the selected fit points; they are not restarted at the lower fit boundary.
We fit $T_H$ independently for each spectrum without an additional normalization parameter.
Since cumulative counts at different points are correlated, the RMSE serves as a measure of descriptive fit quality rather than a statistical significance test.

\begin{table}[t]
\caption{Threshold masses, isospin, strangeness, and total multiplicities of the string channels used in the fits, following Refs.~\cite{Fujimoto:2025trl,Marczenko:2026drt}.
Here, $l$ denotes a light quark ($u$ or $d$), and $[qq]$ and $(qq)$ denote spin-0 and spin-1 diquarks, respectively.
Hadron labels in parentheses identify the reference states used to set the thresholds.
Each $d_i$ includes the endpoint spin multiplicity and all charge states.
Charge-conjugate channels are also included for mesons, while baryons are restricted to $B=1$.}
\label{tab:stringchannels}
\begin{ruledtabular}
\begin{tabular}{lcccc}
Channel $i$ & $I$ & $S$ & $m_i$ [GeV] & $d_i$ \\
\colrule
\multicolumn{5}{l}{Light-meson channels} \\[2pt]
$l\!-\!\bar l$ ($\rho,\omega$) & $1,0$ & 0 & 0.775 & 16 \\
$l\!-\!\bar s$ ($K^*$) & $1/2$ & $\pm1$ & 0.896 & 16 \\
$s\!-\!\bar s$ ($\phi$) & 0 & 0 & 1.019 & 4 \\[5pt]
\multicolumn{5}{l}{Baryon channels ($B=1$)} \\[2pt]
$l\!-\![ll]$ ($N$) & $1/2$ & 0 & 0.938 & 4 \\
$s\!-\![ll]$ ($\Lambda$) & 0 & $-1$ & 1.116 & 2 \\
$l\!-\![ls]$ ($\Sigma$) & 1 & $-1$ & 1.189 & 6 \\
$l\!-\!(ll)$ ($\Delta$) & $3/2$ & 0 & 1.232 & 16 \\
$s\!-\![ls]$ ($\Xi$) & $1/2$ & $-2$ & 1.315 & 4 \\
$s\!-\!(ss)$ ($\Omega$) & 0 & $-3$ & 1.672 & 4 \\[5pt]
\multicolumn{5}{l}{Open-charm channels ($|C|=1$)} \\[2pt]
$c\!-\!\bar l$ ($D$) & $1/2$ & 0 & 1.867 & 16 \\
$c\!-\!\bar s$ ($D_s$) & 0 & $\pm1$ & 1.968 & 8
\end{tabular}
\end{ruledtabular}
\end{table}

\subsection{Bessel spectrum versus the leading term}
\label{sec:besselfit}

Earlier open-string fits based on the leading large-mass density gave $T_H\simeq340\,\mathrm{MeV}$ for both mesons and baryons~\cite{Fujimoto:2025trl}.
We examine the effect of retaining the Bessel function, which incorporates the power corrections to this leading term~\cite{Dienes:1993mp}.
At $b=2$, the Bessel density in Eq.~\eqref{eq:rhoopen} and its leading term are
\begin{align}
    \rho_{\rm open}(m)
    &=\frac{\pi}{3m}I_2\!\left(\frac{m}{T_H}\right)\,,
    \label{eq:rhoopenfit}\\
    \rho_{\rm lead}(m)
    &=\frac{\sqrt{2\pi}}{6T_H}
      \left(\frac{m}{T_H}\right)^{-3/2}e^{m/T_H}\,.
    \label{eq:rhoopenleading}
\end{align}

The accuracy of these approximations can be tested against the exact oscillator multiplicities $p_2(N)$, obtained by expanding the generating function \eqref{eq:Pbw}.
We evaluate both densities at the mass $m_N$ satisfying $\alpha'm_N^2=N-1/12$ and divide by the Jacobian $2\alpha'm_N$ to obtain the corresponding level multiplicities.
Table~\ref{tab:levelaccuracy} shows a substantial improvement with the Bessel form.
At $N=1$, the relative error decreases from $97.1\%$ to $6.40\%$; at $N=10$, it decreases from $18.5\%$ to $0.184\%$.
This improvement in the underlying oscillator counting motivates retaining the Bessel form at the finite masses relevant to hadron spectra.

\begin{table}
\caption{Exact NN oscillator multiplicities $p_2(N)$ and their approximations from the leading and Bessel densities.
Both approximations are evaluated at $\alpha'm_N^2=N-1/12$ and converted to level multiplicities using $p_2(N)\simeq\rho(m_N)/(2\alpha'm_N)$.}
\label{tab:levelaccuracy}
\begin{ruledtabular}
\begin{tabular}{rccc}
$N$ & Exact $p_2(N)$ & Leading term & Bessel \\
\colrule
1  & 2   & 3.94 & 2.13 \\
2  & 5   & 7.38 & 4.89 \\
5  & 36  & 46.6 & 36.4 \\
10 & 481 & 570  & 480
\end{tabular}
\end{ruledtabular}
\end{table}

For the light-meson spectrum, we include the listed light-unflavored and strange mesons with definite spin and a quoted PDG mass, including antiparticles.
Following the threshold prescription of Ref.~\cite{Fujimoto:2025trl}, the cumulative model spectrum is
\begin{align}
    N_M(m) &= 8 + \int_0^m dm'\,d_M(m')\rho(m')\,,
    \label{eq:mesoncumulative}\\
    d_M(m) &= \sum_i d_{M,i} \theta(m - m_{M,i})\,.
\end{align}
The constant $8$ accounts for the Nambu-Goldstone $\pi$, $K$, and $\eta$ states.
The thresholds $m_{M,i}$ and total channel multiplicities $d_{M,i}$ are listed in Table~\ref{tab:stringchannels}.
We use either Eq.~\eqref{eq:rhoopenfit} or Eq.~\eqref{eq:rhoopenleading} for $\rho(m')$.

For baryons, we use the ordinary open-string density in the quark-diquark description~\cite{Fujimoto:2025trl} and construct the cumulative spectrum as
\begin{align}
    N_B(m) &= \int_0^m dm'\,d_B(m')\rho(m')\,,
    \label{eq:baryoncumulative}\\
    d_B(m) &= \sum_i d_{B,i}\theta(m-m_{B,i})\,.
\end{align}
The thresholds $m_{B,i}$ and total channel multiplicities $d_{B,i}$ are listed in Table~\ref{tab:stringchannels}.
These multiplicities are common to the quark-diquark and Y-junction descriptions, which represent the same physical spin-flavor channels within the present modeling.
For the comparison here, we use either Eq.~\eqref{eq:rhoopenfit} or Eq.~\eqref{eq:rhoopenleading} for $\rho$.
The PDG input is restricted to established three- and four-star $u,d,s$ baryons with $B=1$.

For both sectors, states at identical masses are combined before constructing the cumulative count.
The strict fit window $1\,\mathrm{GeV}<m<2\,\mathrm{GeV}$ gives $n_{\rm fit}=66$ points for mesons and $47$ for baryons.

\begin{figure}
    \centering
    \includegraphics[width=0.95\columnwidth]{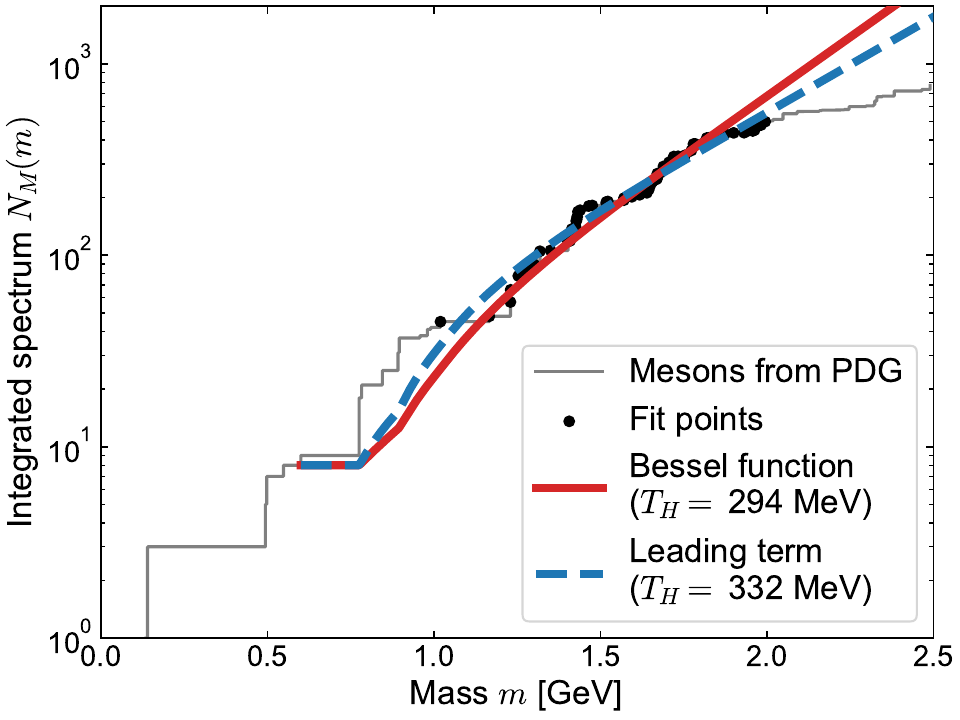}
    \includegraphics[width=0.95\columnwidth]{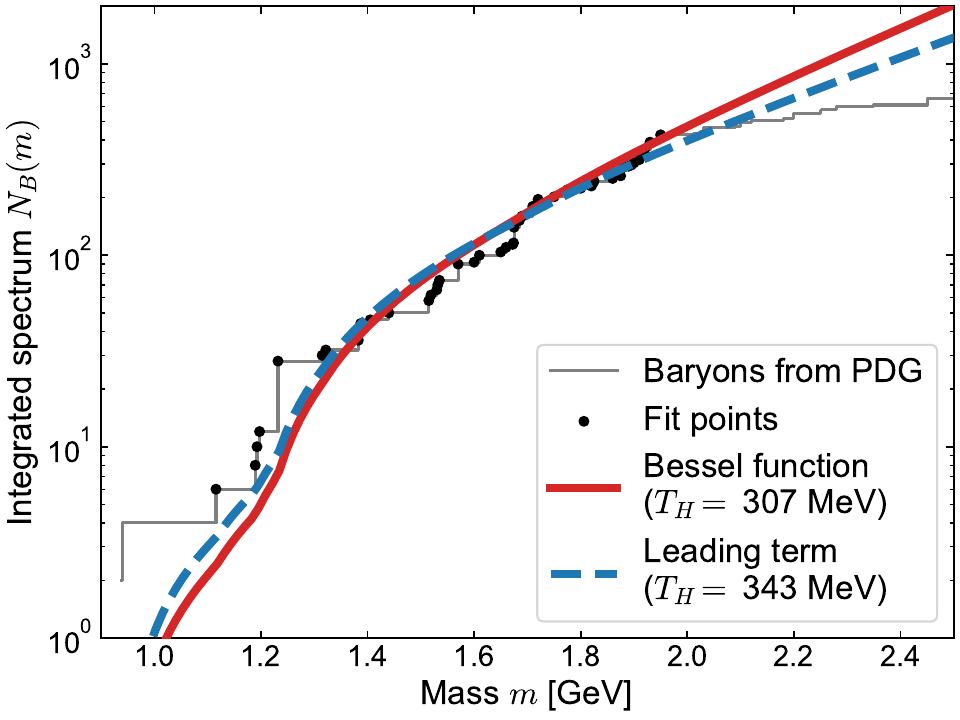}
    \caption{Cumulative light-meson (top) and quark-diquark baryon (bottom) spectra obtained with the Bessel density (red solid curves) and its leading term (blue dashed curves).
    The grey lines show the PDG 2024 cumulative spectra, and the black points mark the $66$ meson and $47$ baryon fit points in $1\,\mathrm{GeV}<m<2\,\mathrm{GeV}$.
    The thresholds and absolute normalizations are common to the two approximations.}
    \label{fig:leadingbessel}
\end{figure}

\begin{table}
\caption{Temperatures and logarithmic RMSE obtained with the leading and Bessel densities, using $66$ meson and $47$ baryon cumulative points in $1\,\mathrm{GeV}<m<2\,\mathrm{GeV}$.}
\label{tab:leadingbessel}
\begin{ruledtabular}
\begin{tabular}{llcc}
Sector & Density & $T_H$ [MeV] & log RMSE \\
\colrule
\multirow{2}{*}{Meson}
& Leading term & 332 & 0.109 \\
& Bessel       & 294 & 0.154 \\[5pt]
\multirow{2}{*}{Baryon}
& Leading term & 343 & 0.268 \\
& Bessel       & 307 & 0.345
\end{tabular}
\end{ruledtabular}
\end{table}

The results are shown in Fig.~\ref{fig:leadingbessel} and Table~\ref{tab:leadingbessel}.
The leading-term fits give $T_H\simeq332\,\mathrm{MeV}$ for mesons and $343\,\mathrm{MeV}$ for baryons, close to the scale reported in Ref.~\cite{Fujimoto:2025trl}.
Retaining the Bessel function lowers these values to $294\,\mathrm{MeV}$ and $307\,\mathrm{MeV}$, respectively.
At a fixed $T_H$, the Bessel density is smaller than the leading term over the fitted range, and a lower temperature compensates for this suppression.
The Bessel fits have slightly larger logarithmic RMSE, but their improved accuracy in the oscillator counting provides a theoretical basis for this change in the inferred temperature.
The resulting common scale $T_H\simeq300\,\mathrm{MeV}$ will serve as a reference for comparing the string configurations below.

\subsection{Open charm spectra: fixed endpoints versus open strings}
\label{sec:charmfit}

The large mass difference between the charm quark and the light antiquark motivates treating an open-charm meson as a string with one fixed endpoint.
We compare this prescription with an ordinary open string with two free endpoints.
The two spectra share the same Hagedorn temperature at a common string tension, but their different oscillator spectra give different Bessel functions and large-energy prefactors.
Setting $b=2$ in Eq.~\eqref{eq:rhoopen} gives
\begin{align}
    \rho_{\rm open}(\varepsilon) =
    \frac{\pi}{3\varepsilon}I_2\!\left(\frac{\varepsilon}{T_H}\right)\,.
    \label{eq:rhoopen_e}
\end{align}
Here, we define the string energy $\varepsilon$ as
\begin{align}
    \varepsilon(m) = m - \sum_f n_f m_f = m - m_c - n_s m_s\,,
    \label{eq:strene}
\end{align}
where $n_f$ is the number of quarks or antiquarks of flavor $f$ with mass $m_f$, and $n_s=0$ or $1$ for the nonstrange and strange open-charm channels, respectively.
The $u$ and $d$ quarks are treated as massless.
We follow the approach taken in Ref.~\cite{Marczenko:2026drt} and subtract the charm quark mass as well as the strange quark mass from the total mass.
The corresponding fixed-endpoint spectrum follows from Eq.~\eqref{eq:rhofixed},
\begin{align}
    \rho_{\rm fixed}(\varepsilon) =
    \frac{1}{2T_H}I_1\!\left(\frac{\varepsilon}{T_H}\right)\,.
    \label{eq:rhofixed_e}
\end{align}

We consider the cumulative mass spectrum for open charm mesons:
\begin{align}
    N_C(m) &= \sum_i d_{C,i}\theta(m-m_{C,i}) \notag\\
    &\quad \times \int_{m_{C,i}}^m dm'\,
    \rho\big(m'-m_c-n_{s,i}m_s\big)\,.
    \label{eq:charmcumulative}
\end{align}
Here, $i$ labels the string channels, and $n_{s,i}$ specifies their strange-quark content.
We take either Eq.~\eqref{eq:rhoopen_e} or Eq.~\eqref{eq:rhofixed_e} for $\rho$.
The total channel multiplicities $d_{C,i}$ and physical thresholds $m_{C,i}$ are listed in Table~\ref{tab:stringchannels}.

For the fit, we order the PDG states by the mass with the strange-quark contribution subtracted, $m_*=m-n_s m_s$, with $m_s = 93.5\,\mathrm{MeV}$, and form the cumulative count in this variable.
We select $n_{\rm fit} = 16$ cumulative points with $m_* - m_c^{\rm cur}>1\,\mathrm{GeV}$, where the charm quark current-mass benchmark is $m_c^{\rm cur}\simeq1.27\,\mathrm{GeV}$.
This cut is made once and kept fixed when $m_c$ is varied later.
The figures display the cumulative spectra ordered by the physical hadron mass $m$, with the model curves given by Eq.~\eqref{eq:charmcumulative}.

We perform two fits for each density: one with $m_c = 1.27\,\mathrm{GeV}$ fixed and $T_H$ fitted, and one with both $T_H$ and $m_c$ fitted.
Here, $m_c$ is an effective endpoint mass and need not coincide with the current quark mass.
In the latter case, we restrict $m_c$ to $[1.27\,\mathrm{GeV},M_{J/\psi}/2]$, with $M_{J/\psi} = 3.10\,\text{GeV}$ to explore the dependence on the endpoint mass between the current-mass benchmark and a constituent-mass estimate.
The physical thresholds and channel degeneracies remain fixed in both fits.

\begin{figure}
    \centering
    \includegraphics[width=0.95\columnwidth]{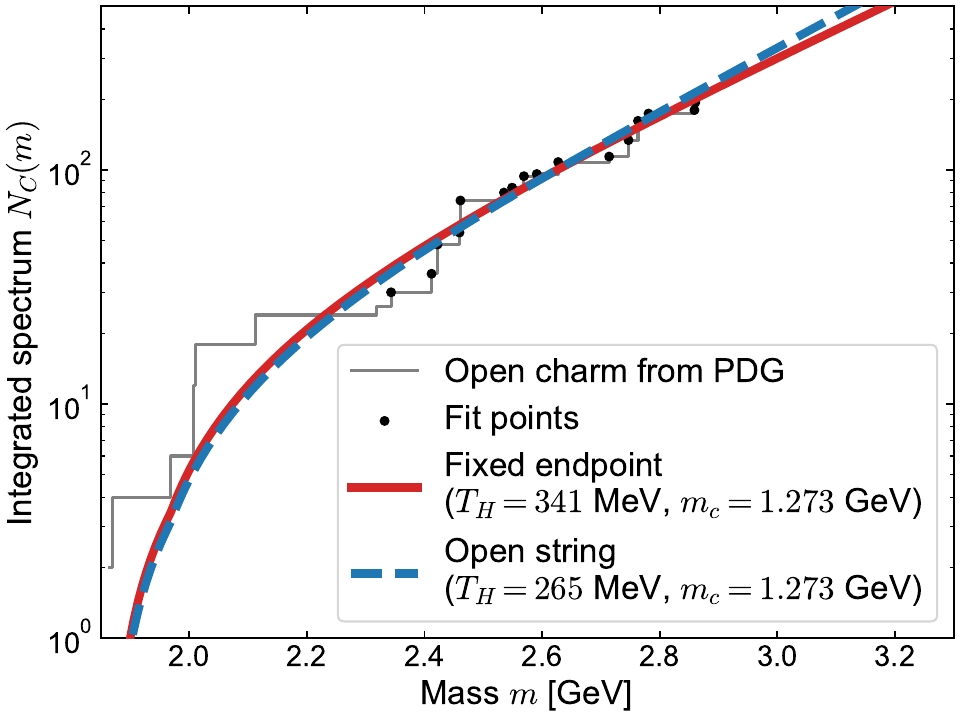}
    \caption{The cumulative open-charm meson spectrum from PDG (grey line) and the curves obtained with the fixed-endpoint Bessel density (red solid curve) and the open-string Bessel density (blue dashed curve), with $m_c=1.273\,\mathrm{GeV}$ fixed.
    The points mark the selected states; the fit is performed using the cumulative count ordered by $m_*=m-n_s m_s$, as described in the text.}
    \label{fig:mcfixed}
\end{figure}

\begin{figure}
    \centering
    \includegraphics[width=0.95\columnwidth]{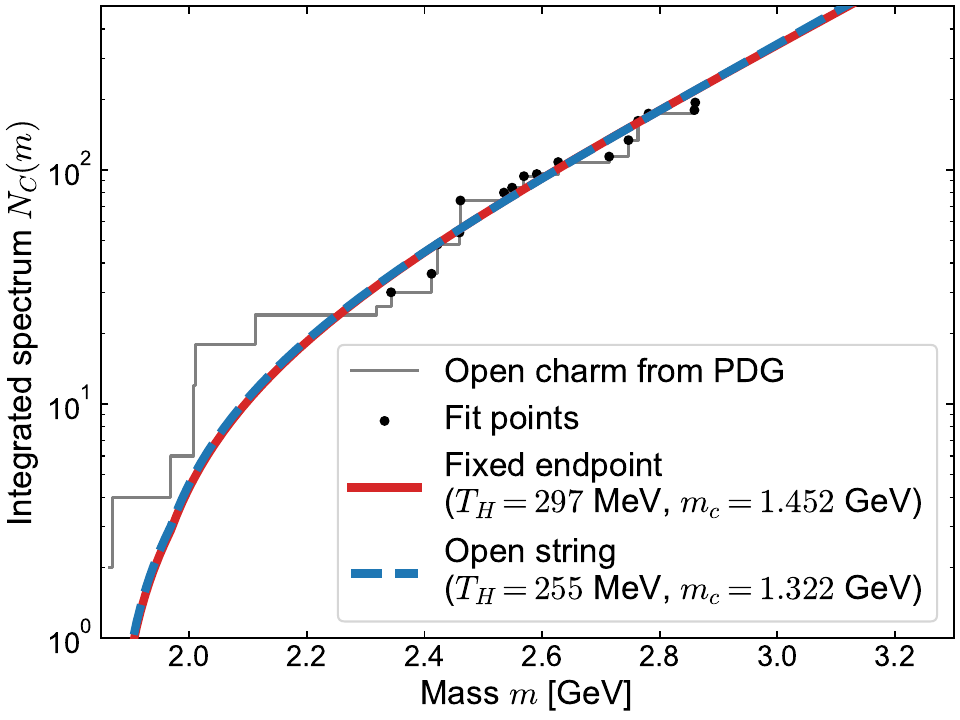}
    \caption{Same comparison as in Fig.~\ref{fig:mcfixed}, with the charm quark mass in Eq.~\eqref{eq:strene} also varied as a fit parameter.}
    \label{fig:mcfitted}
\end{figure}

\begin{table}
\caption{Fitted parameters and logarithmic RMSE for the open-charm cumulative spectrum using the Bessel densities.
The fixed-$m_c$ and fitted-$m_c$ results correspond to Figs.~\ref{fig:mcfixed} and \ref{fig:mcfitted}, respectively.}
\label{tab:charm}
\begin{ruledtabular}
\begin{tabular}{llccc}
Condition
& Model
& $T_H$ [MeV]
& $m_c$ [GeV]
& log RMSE \\
\hline
 & & & & \\[-8pt]
\multirow{2}{*}{$m_c$ fixed}
& Open string
& 265
& 1.27
& 0.172 \\
& Fixed endpoint
& 341
& 1.27
& 0.177 \\[5pt]
\multirow{2}{*}{$m_c$ fitted}
& Open string
& 255
& 1.32
& 0.172 \\
& Fixed endpoint
& 297
& 1.45
& 0.171
\end{tabular}
\end{ruledtabular}
\end{table}

The results are shown in Figs.~\ref{fig:mcfixed} and \ref{fig:mcfitted} and summarized in Table~\ref{tab:charm}.
With $m_c$ fixed to the current-mass benchmark, the open-string and fixed-endpoint fits give $T_H\simeq265\,\mathrm{MeV}$ and $341\,\mathrm{MeV}$, respectively.
When the effective endpoint mass is fitted, the open-string spectrum gives $T_H\simeq255\,\mathrm{MeV}$ with $m_c\simeq1.32\,\mathrm{GeV}$, whereas the fixed-endpoint spectrum gives $T_H\simeq297\,\mathrm{MeV}$ with $m_c\simeq1.45\,\mathrm{GeV}$.
The two fitted-mass curves have nearly identical logarithmic RMSE, but the fixed-endpoint result agrees with the $T_H\simeq300\,\mathrm{MeV}$ scale extracted from light mesons and baryons in Sec.~\ref{sec:besselfit}.
This consistency across sectors supports the fixed-endpoint description of the heavy-light string.

\subsection{Baryon spectra: junction versus quark-diquark descriptions}
\label{sec:baryonfit}

For baryons, we compare the Y-junction spectrum with the ordinary open-string spectrum used in the quark-diquark description~\cite{Fujimoto:2025trl}.
At $b=2$, the Bessel densities in Eqs.~\eqref{eq:rhoopen} and \eqref{eq:rhoY} are
\begin{align}
    \rho_{\rm open}(m)
    &=
    \frac{\pi}{3m}I_2\!\left(\frac{m}{T_H}\right)\,,\\
    \rho_{Y}(m) &= \frac34 \rho_{\rm open}(m)\,.
    \label{eq:baryondensities}
\end{align}
Unlike the open-charm case, the two densities have the same Bessel function and mass-dependent prefactor.
Their only difference is the overall factor $3/4$.
We use the cumulative spectrum in Eq.~\eqref{eq:baryoncumulative} with the same channels, PDG input, and fit window as in Sec.~\ref{sec:besselfit}, replacing $\rho_{\rm open}$ by $\rho_Y$ for the junction fit.
Both descriptions therefore use the same $47$ cumulative points.

\begin{figure}
    \centering
    \includegraphics[width=0.95\columnwidth]{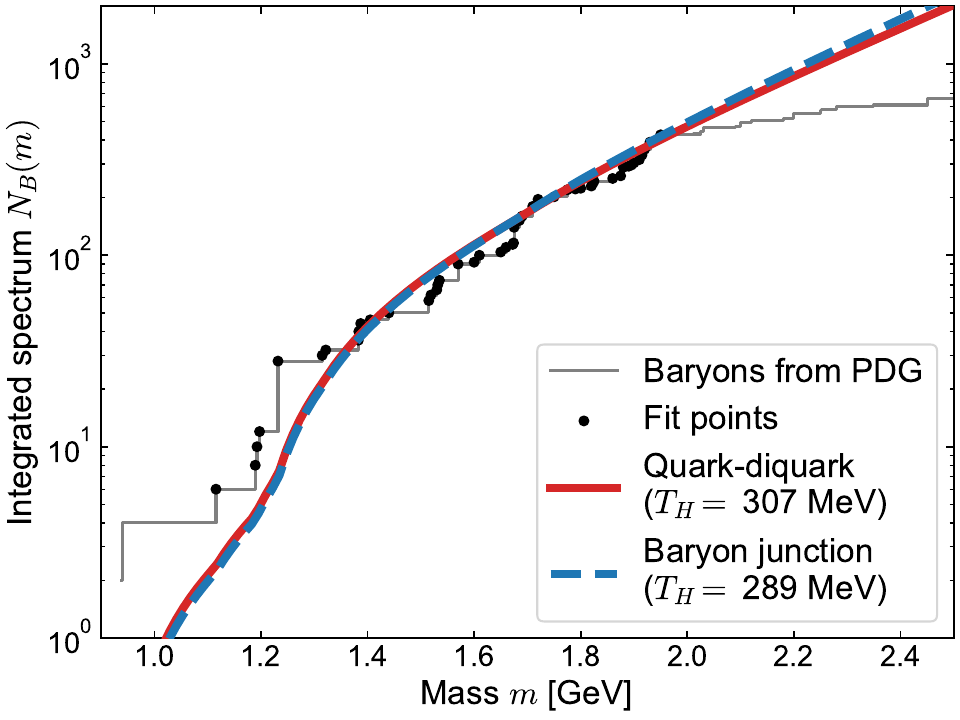}
    \caption{The cumulative baryon mass spectrum from PDG (grey line) and the curves fitted with the open-string Bessel density for a quark-diquark configuration (red solid curve) and the Y-junction Bessel density (blue dashed curve).
    The points mark the cumulative counts used in the fit.}
    \label{fig:junction}
\end{figure}

Figure~\ref{fig:junction} shows that both descriptions reproduce the overall growth of the cumulative baryon spectrum with comparable accuracy in the fitted mass range.
The quark-diquark fit gives $T_H\simeq307\,\mathrm{MeV}$ and $\mathrm{RMSE}_{\ln N}=0.345$, while the junction fit gives $T_H\simeq289\,\mathrm{MeV}$ and $\mathrm{RMSE}_{\ln N}=0.367$.
Compared with the leading-term results, $343\,\mathrm{MeV}$ and $319\,\mathrm{MeV}$, the Bessel fits bring both temperatures close to the $300\,\mathrm{MeV}$ scale.
The smaller normalization of the junction density is largely compensated by a lower fitted $T_H$, which enhances the growth over this finite mass interval.
Consequently, at the hadron masses available here, the cumulative mass spectrum provides little discrimination between the two descriptions.

The comparison with Regge trajectories introduces an additional constraint.
From Eq.~\eqref{eq:THopen}, the fitted temperature determines the string parameter through
\begin{align}
    \alpha'=\frac{3}{4\pi^2 T_H^2}\,.
\end{align}
The classical relations, as explored in Sec.~\ref{sec:Regge}, give
\begin{align}
    \left.\frac{dJ}{dm^2}\right|_{\rm quark\text{-}diquark}
    &=\alpha'\,,\qquad
    \left.\frac{dJ}{dm^2}\right|_{Y}
    =\frac23\alpha'\,.
    \label{eq:baryonreggeslopes}
\end{align}
Using the respective Bessel-fit temperatures, these slopes are approximately $0.809\,\mathrm{GeV}^{-2}$ and $0.606\,\mathrm{GeV}^{-2}$.
Prominent light-baryon Regge trajectories instead have slopes close to those of light mesons, with a characteristic value around $0.9\,\mathrm{GeV}^{-2}$~\cite{Selem:2006nd}.
For comparison, the light-meson Bessel fit in Sec.~\ref{sec:besselfit} corresponds to $\alpha'\simeq0.879\,\mathrm{GeV}^{-2}$.
Retaining the Bessel form therefore improves the consistency of the quark-diquark temperature with the observed Regge slope.
Thus, although the two cumulative mass spectra are difficult to distinguish, the Y-junction description has a larger tension with the observed Regge trajectories.
At a common string tension, it predicts a slope equal to two thirds of the open-string slope, and the shift in $T_H$ obtained from the spectrum fit does not compensate for this reduction.

\subsection{Thermodynamics of the Bessel spectrum}
\label{sec:thermo}

The Bessel spectrum also gives a simple analytic expression for the pressure of an ideal string gas.
We consider a single channel with unit degeneracy and integrate the continuum density from $m = 0$.
The Maxwell-Boltzmann pressure of a mass $m$ state is
\begin{align}
    P_{\rm MB}(T,m) = \frac{m^2T^2}{2\pi^2}K_2\!\left(\frac{m}{T}\right)\,.
    \label{eq:boltzmannpressure}
\end{align}
Using the spectrum in Eq.~\eqref{eq:rhoopenfit} and a standard Bessel integral, we obtain
\begin{align}
    \frac{P_{\rm open}(T)}{T^4}
    &= \frac1{T^4}\int_0^\infty dm\,\rho_{\rm open}(m)P_{\rm MB}(T,m)
    \notag\\
    &= \frac{T^2}{6\pi(T_H^2-T^2)}\,.
    \label{eq:besselpressure}
\end{align}
The condition $T<T_H$ ensures convergence at large mass.

Equation~\eqref{eq:besselpressure} displays a simple pole at the Hagedorn temperature.
For the junction density in Eq.~\eqref{eq:baryondensities}, the same calculation gives
\begin{align}
    P_Y(T)=\frac34 P_{\rm open}(T)\,.
    \label{eq:junctionpressure}
\end{align}
The junction normalization therefore changes the magnitude of the pressure while preserving the limiting temperature.
The same factor relates the ideal-gas entropy and energy densities at a common $T_H$.

The rational form in Eq.~\eqref{eq:besselpressure} assumes a continuum extending to zero mass.
Hadronic thermodynamics also requires the finite channel thresholds, multiplicities, and discrete low-lying states.
A finite lower threshold subtracts a contribution that remains finite at $T=T_H$, leaving the pole and its residue unchanged.
Applications to the hadron resonance gas are beyond the scope of this work and will be reported elsewhere.

\section{Conclusions}

We have derived the mass formulas for strings with one fixed heavy endpoint and for three equal-tension strings meeting at a massless baryon junction.
Their asymptotic densities of states are expressed in terms of modified Bessel functions, retaining power corrections to the leading large-mass terms.
Within our prescription of the state counting for noninteracting oscillators, both share the Hagedorn temperature of the ordinary open string when the string tension and number of polarizations are the same.
In three spatial dimensions, the common temperature is $T_H=\sqrt{3\mathcal{T}/(2\pi)}$, which also applies to closed strings used to model glueballs~\cite{Meyer:2009tq,Fujimoto:2025sxx}.
This result provides a microscopic explanation for the common Hagedorn temperature used in string descriptions of light hadrons and heavy flavors.

The universality of $T_H$ is compatible with different leading Regge slopes.
Regge trajectories select maximum-spin configurations governed by the lowest circularly polarized mode, whereas the leading growth of the total state count is controlled by the asymptotic density of oscillator frequencies.
Fixing one endpoint shifts the frequencies without changing their asymptotic density.
For the junction, the additional oscillator towers and the normalization of the total string energy involve the same sum over arms, which cancels in the leading exponential growth per unit mass.
The corresponding mass formulas are therefore consistent with the classical Regge relations without requiring different Hagedorn temperatures.

The Bessel expressions substantially improve the agreement with the exact NN oscillator multiplicities $p_2(N)$ compared with the leading large-mass term.
In fits to light mesons and quark-diquark baryons, retaining the Bessel form lowers $T_H$ from $332\,\mathrm{MeV}$ and $343\,\mathrm{MeV}$ to $294\,\mathrm{MeV}$ and $307\,\mathrm{MeV}$, respectively.
The improved oscillator counting thus brings both fitted temperatures close to the common scale of $300\,\mathrm{MeV}$.

The spectral prefactors retain information about the boundary and junction conditions.
With the effective charm endpoint mass fitted, the fixed-endpoint density gives $T_H\simeq297\,\mathrm{MeV}$, consistent with this scale, whereas the ordinary open-string fit gives $T_H\simeq255\,\mathrm{MeV}$ with comparable RMSE.
The baryon-junction density is $3/4$ of the quark-diquark density at the same $T_H$.
When $T_H$ is fitted independently, this normalization difference is largely compensated by a shift in temperature, and the available baryon spectrum does not clearly favor either picture.
Regge trajectories provide an additional constraint: the Y junction predicts a slope of $2\alpha'/3$, and the fitted temperatures yield a larger discrepancy with the observed light-baryon trajectories than in the quark-diquark description.

For an ideal Boltzmann gas with a continuum extending to zero mass, the Bessel spectrum also yields a rational expression for the pressure with a simple pole at $T=T_H$.
At a common $T_H$, the junction pressure is $3/4$ of the open-string pressure.

These phenomenological comparisons use asymptotic densities at finite hadron masses and depend on the endpoint mass and channel-threshold prescriptions.
Including finite endpoint and junction masses and corrections beyond the independent-oscillator approximation would help test the descriptions against both cumulative state counts and individual Regge trajectories.
The present results establish how a baryon-junction spectrum can reproduce the same Hagedorn growth as an open string while retaining differences in normalization and angular-momentum dependence.

The common Hagedorn temperature also has implications for deconfinement.
Following the interpretation of Cabibbo and Parisi, it characterizes the limiting scale of the confined string description and can be viewed as an upper scale for quark and gluon liberation~\cite{Cabibbo:1975ig}.
Its value of order 300 MeV, shared by the meson, baryon, and glueball string descriptions, is close to the pure SU(3) Yang-Mills deconfinement temperature, $T_d \simeq 285 \,\mathrm{MeV}$~\cite{Borsanyi:2022xml,Giusti:2025fxu,Svetitsky:1985ye}.
This suggests a common upper scale for deconfinement across these sectors.

\begin{acknowledgments}
I thank Michal Marczenko for correspondence and for drawing my attention to Ref.~\cite{Dienes:1993mp}.
This work is supported by JSPS KAKENHI Nos.~25K23388 and 26K17140.
\end{acknowledgments}

\bibliography{string}

\begin{thebibliography}{57}%
\makeatletter
\providecommand \@ifxundefined [1]{%
 \@ifx{#1\undefined}
}%
\providecommand \@ifnum [1]{%
 \ifnum #1\expandafter \@firstoftwo
 \else \expandafter \@secondoftwo
 \fi
}%
\providecommand \@ifx [1]{%
 \ifx #1\expandafter \@firstoftwo
 \else \expandafter \@secondoftwo
 \fi
}%
\providecommand \natexlab [1]{#1}%
\providecommand \enquote  [1]{``#1''}%
\providecommand \bibnamefont  [1]{#1}%
\providecommand \bibfnamefont [1]{#1}%
\providecommand \citenamefont [1]{#1}%
\providecommand \href@noop [0]{\@secondoftwo}%
\providecommand \href [0]{\begingroup \@sanitize@url \@href}%
\providecommand \@href[1]{\@@startlink{#1}\@@href}%
\providecommand \@@href[1]{\endgroup#1\@@endlink}%
\providecommand \@sanitize@url [0]{\catcode `\\12\catcode `\$12\catcode `\&12\catcode `\#12\catcode `\^12\catcode `\_12\catcode `\%12\relax}%
\providecommand \@@startlink[1]{}%
\providecommand \@@endlink[0]{}%
\providecommand \url  [0]{\begingroup\@sanitize@url \@url }%
\providecommand \@url [1]{\endgroup\@href {#1}{\urlprefix }}%
\providecommand \urlprefix  [0]{URL }%
\providecommand \Eprint [0]{\href }%
\providecommand \doibase [0]{https://doi.org/}%
\providecommand \selectlanguage [0]{\@gobble}%
\providecommand \bibinfo  [0]{\@secondoftwo}%
\providecommand \bibfield  [0]{\@secondoftwo}%
\providecommand \translation [1]{[#1]}%
\providecommand \BibitemOpen [0]{}%
\providecommand \bibitemStop [0]{}%
\providecommand \bibitemNoStop [0]{.\EOS\space}%
\providecommand \EOS [0]{\spacefactor3000\relax}%
\providecommand \BibitemShut  [1]{\csname bibitem#1\endcsname}%
\let\auto@bib@innerbib\@empty
\bibitem [{\citenamefont {Fujimoto}\ \emph {et~al.}(2025)\citenamefont {Fujimoto}, \citenamefont {Fukushima}, \citenamefont {Hidaka},\ and\ \citenamefont {McLerran}}]{Fujimoto:2025sxx}%
  \BibitemOpen
  \bibfield  {author} {\bibinfo {author} {\bibfnamefont {Y.}~\bibnamefont {Fujimoto}}, \bibinfo {author} {\bibfnamefont {K.}~\bibnamefont {Fukushima}}, \bibinfo {author} {\bibfnamefont {Y.}~\bibnamefont {Hidaka}},\ and\ \bibinfo {author} {\bibfnamefont {L.}~\bibnamefont {McLerran}},\ }\bibfield  {title} {\bibinfo {title} {{New state of matter between the hadronic phase and the quark-gluon plasma?}},\ }\href {https://doi.org/10.1103/h71y-km92} {\bibfield  {journal} {\bibinfo  {journal} {Phys. Rev. D}\ }\textbf {\bibinfo {volume} {112}},\ \bibinfo {pages} {074006} (\bibinfo {year} {2025})},\ \Eprint {https://arxiv.org/abs/2506.00237} {arXiv:2506.00237 [hep-ph]} \BibitemShut {NoStop}%
\bibitem [{\citenamefont {Marczenko}\ \emph {et~al.}(2025)\citenamefont {Marczenko}, \citenamefont {Kov{\'a}cs}, \citenamefont {McLerran},\ and\ \citenamefont {Redlich}}]{Marczenko:2025nhj}%
  \BibitemOpen
  \bibfield  {author} {\bibinfo {author} {\bibfnamefont {M.}~\bibnamefont {Marczenko}}, \bibinfo {author} {\bibfnamefont {G.}~\bibnamefont {Kov{\'a}cs}}, \bibinfo {author} {\bibfnamefont {L.}~\bibnamefont {McLerran}},\ and\ \bibinfo {author} {\bibfnamefont {K.}~\bibnamefont {Redlich}},\ }\bibfield  {title} {\bibinfo {title} {{String-based model with Hagedorn temperature of TH{\ensuremath{\sim}}300{\,}{\,}MeV describes the spectrum of mesons and glueballs}},\ }\href {https://doi.org/10.1103/cgyz-3956} {\bibfield  {journal} {\bibinfo  {journal} {Phys. Rev. D}\ }\textbf {\bibinfo {volume} {112}},\ \bibinfo {pages} {096010} (\bibinfo {year} {2025})},\ \Eprint {https://arxiv.org/abs/2508.11626} {arXiv:2508.11626 [hep-ph]} \BibitemShut {NoStop}%
\bibitem [{\citenamefont {Fujimoto}(2026)}]{Fujimoto:2025trl}%
  \BibitemOpen
  \bibfield  {author} {\bibinfo {author} {\bibfnamefont {Y.}~\bibnamefont {Fujimoto}},\ }\bibfield  {title} {\bibinfo {title} {{Description of the baryon mass spectrum by open strings and diquarks}},\ }\href {https://doi.org/10.1103/n6b5-f2t3} {\bibfield  {journal} {\bibinfo  {journal} {Phys. Rev. D}\ }\textbf {\bibinfo {volume} {113}},\ \bibinfo {pages} {054044} (\bibinfo {year} {2026})},\ \Eprint {https://arxiv.org/abs/2512.24882} {arXiv:2512.24882 [hep-ph]} \BibitemShut {NoStop}%
\bibitem [{\citenamefont {Marczenko}\ and\ \citenamefont {Redlich}(2026)}]{Marczenko:2026yme}%
  \BibitemOpen
  \bibfield  {author} {\bibinfo {author} {\bibfnamefont {M.}~\bibnamefont {Marczenko}}\ and\ \bibinfo {author} {\bibfnamefont {K.}~\bibnamefont {Redlich}},\ }\bibfield  {title} {\bibinfo {title} {{Evidence for quark-diquark structure of baryons from fluctuations of conserved charges}},\ }\href {https://doi.org/10.1103/yb26-gj61} {\bibfield  {journal} {\bibinfo  {journal} {Phys. Rev. C}\ }\textbf {\bibinfo {volume} {114}},\ \bibinfo {pages} {025205} (\bibinfo {year} {2026})},\ \Eprint {https://arxiv.org/abs/2601.22902} {arXiv:2601.22902 [hep-ph]} \BibitemShut {NoStop}%
\bibitem [{\citenamefont {Marczenko}\ \emph {et~al.}(2026)\citenamefont {Marczenko}, \citenamefont {McLerran},\ and\ \citenamefont {Redlich}}]{Marczenko:2026drt}%
  \BibitemOpen
  \bibfield  {author} {\bibinfo {author} {\bibfnamefont {M.}~\bibnamefont {Marczenko}}, \bibinfo {author} {\bibfnamefont {L.}~\bibnamefont {McLerran}},\ and\ \bibinfo {author} {\bibfnamefont {K.}~\bibnamefont {Redlich}},\ }\bibfield  {title} {\bibinfo {title} {{Hadron spectra and thermodynamics for all quark flavors from a universal Hagedorn temperature}},\ }\href@noop {} {\  (\bibinfo {year} {2026})},\ \Eprint {https://arxiv.org/abs/2603.28668} {arXiv:2603.28668 [hep-ph]} \BibitemShut {NoStop}%
\bibitem [{\citenamefont {Fujimoto}\ and\ \citenamefont {Vovchenko}(2026)}]{Fujimoto:2026syp}%
  \BibitemOpen
  \bibfield  {author} {\bibinfo {author} {\bibfnamefont {Y.}~\bibnamefont {Fujimoto}}\ and\ \bibinfo {author} {\bibfnamefont {V.}~\bibnamefont {Vovchenko}},\ }\bibfield  {title} {\bibinfo {title} {{QCD thermodynamics through the crossover as a gas of confining strings with repulsive interactions}},\ }\href@noop {} {\  (\bibinfo {year} {2026})},\ \Eprint {https://arxiv.org/abs/2609.23292} {arXiv:2609.23292 [hep-ph]} \BibitemShut {NoStop}%
\bibitem [{\citenamefont {Rohrhofer}\ \emph {et~al.}(2019)\citenamefont {Rohrhofer}, \citenamefont {Aoki}, \citenamefont {Cossu}, \citenamefont {Fukaya}, \citenamefont {Gattringer}, \citenamefont {Glozman}, \citenamefont {Hashimoto}, \citenamefont {Lang},\ and\ \citenamefont {Prelovsek}}]{Rohrhofer:2019qwq}%
  \BibitemOpen
  \bibfield  {author} {\bibinfo {author} {\bibfnamefont {C.}~\bibnamefont {Rohrhofer}}, \bibinfo {author} {\bibfnamefont {Y.}~\bibnamefont {Aoki}}, \bibinfo {author} {\bibfnamefont {G.}~\bibnamefont {Cossu}}, \bibinfo {author} {\bibfnamefont {H.}~\bibnamefont {Fukaya}}, \bibinfo {author} {\bibfnamefont {C.}~\bibnamefont {Gattringer}}, \bibinfo {author} {\bibfnamefont {L.~Y.}\ \bibnamefont {Glozman}}, \bibinfo {author} {\bibfnamefont {S.}~\bibnamefont {Hashimoto}}, \bibinfo {author} {\bibfnamefont {C.~B.}\ \bibnamefont {Lang}},\ and\ \bibinfo {author} {\bibfnamefont {S.}~\bibnamefont {Prelovsek}},\ }\bibfield  {title} {\bibinfo {title} {{Symmetries of spatial meson correlators in high temperature QCD}},\ }\href {https://doi.org/10.1103/PhysRevD.100.014502} {\bibfield  {journal} {\bibinfo  {journal} {Phys. Rev. D}\ }\textbf {\bibinfo {volume} {100}},\ \bibinfo {pages} {014502} (\bibinfo {year} {2019})},\ \Eprint {https://arxiv.org/abs/1902.03191} {arXiv:1902.03191 [hep-lat]} \BibitemShut {NoStop}%
\bibitem [{\citenamefont {Rohrhofer}\ \emph {et~al.}(2020)\citenamefont {Rohrhofer}, \citenamefont {Aoki}, \citenamefont {Glozman},\ and\ \citenamefont {Hashimoto}}]{Rohrhofer:2019qal}%
  \BibitemOpen
  \bibfield  {author} {\bibinfo {author} {\bibfnamefont {C.}~\bibnamefont {Rohrhofer}}, \bibinfo {author} {\bibfnamefont {Y.}~\bibnamefont {Aoki}}, \bibinfo {author} {\bibfnamefont {L.~Y.}\ \bibnamefont {Glozman}},\ and\ \bibinfo {author} {\bibfnamefont {S.}~\bibnamefont {Hashimoto}},\ }\bibfield  {title} {\bibinfo {title} {{Chiral-spin symmetry of the meson spectral function above $T_c$}},\ }\href {https://doi.org/10.1016/j.physletb.2020.135245} {\bibfield  {journal} {\bibinfo  {journal} {Phys. Lett. B}\ }\textbf {\bibinfo {volume} {802}},\ \bibinfo {pages} {135245} (\bibinfo {year} {2020})},\ \Eprint {https://arxiv.org/abs/1909.00927} {arXiv:1909.00927 [hep-lat]} \BibitemShut {NoStop}%
\bibitem [{\citenamefont {Glozman}\ \emph {et~al.}(2022)\citenamefont {Glozman}, \citenamefont {Philipsen},\ and\ \citenamefont {Pisarski}}]{Glozman:2022lda}%
  \BibitemOpen
  \bibfield  {author} {\bibinfo {author} {\bibfnamefont {L.~Y.}\ \bibnamefont {Glozman}}, \bibinfo {author} {\bibfnamefont {O.}~\bibnamefont {Philipsen}},\ and\ \bibinfo {author} {\bibfnamefont {R.~D.}\ \bibnamefont {Pisarski}},\ }\bibfield  {title} {\bibinfo {title} {{Chiral spin symmetry and the QCD phase diagram}},\ }\href {https://doi.org/10.1140/epja/s10050-022-00895-4} {\bibfield  {journal} {\bibinfo  {journal} {Eur. Phys. J. A}\ }\textbf {\bibinfo {volume} {58}},\ \bibinfo {pages} {247} (\bibinfo {year} {2022})},\ \Eprint {https://arxiv.org/abs/2204.05083} {arXiv:2204.05083 [hep-ph]} \BibitemShut {NoStop}%
\bibitem [{\citenamefont {Cohen}\ and\ \citenamefont {Glozman}(2024)}]{Cohen:2023hbq}%
  \BibitemOpen
  \bibfield  {author} {\bibinfo {author} {\bibfnamefont {T.~D.}\ \bibnamefont {Cohen}}\ and\ \bibinfo {author} {\bibfnamefont {L.~Y.}\ \bibnamefont {Glozman}},\ }\bibfield  {title} {\bibinfo {title} {{Large $N_c$ QCD phase diagram at $\mu _B=0$}},\ }\href {https://doi.org/10.1140/epja/s10050-024-01400-9} {\bibfield  {journal} {\bibinfo  {journal} {Eur. Phys. J. A}\ }\textbf {\bibinfo {volume} {60}},\ \bibinfo {pages} {171} (\bibinfo {year} {2024})},\ \Eprint {https://arxiv.org/abs/2311.07333} {arXiv:2311.07333 [hep-ph]} \BibitemShut {NoStop}%
\bibitem [{\citenamefont {Hanada}\ \emph {et~al.}(2026)\citenamefont {Hanada}, \citenamefont {Holden},\ and\ \citenamefont {Watanabe}}]{Hanada:2025rca}%
  \BibitemOpen
  \bibfield  {author} {\bibinfo {author} {\bibfnamefont {M.}~\bibnamefont {Hanada}}, \bibinfo {author} {\bibfnamefont {J.}~\bibnamefont {Holden}},\ and\ \bibinfo {author} {\bibfnamefont {H.}~\bibnamefont {Watanabe}},\ }\bibfield  {title} {\bibinfo {title} {{New Phases in QCD at Finite Temperature and Chemical Potential}},\ }\href {https://doi.org/10.1093/ptep/ptag124} {\bibfield  {journal} {\bibinfo  {journal} {PTEP}\ }\textbf {\bibinfo {volume} {2026}},\ \bibinfo {pages} {083B03} (\bibinfo {year} {2026})},\ \Eprint {https://arxiv.org/abs/2509.04671} {arXiv:2509.04671 [hep-th]} \BibitemShut {NoStop}%
\bibitem [{\citenamefont {Veneziano}(1968)}]{Veneziano:1968yb}%
  \BibitemOpen
  \bibfield  {author} {\bibinfo {author} {\bibfnamefont {G.}~\bibnamefont {Veneziano}},\ }\bibfield  {title} {\bibinfo {title} {{Construction of a crossing - symmetric, Regge behaved amplitude for linearly rising trajectories}},\ }\href {https://doi.org/10.1007/BF02824451} {\bibfield  {journal} {\bibinfo  {journal} {Nuovo Cim. A}\ }\textbf {\bibinfo {volume} {57}},\ \bibinfo {pages} {190} (\bibinfo {year} {1968})}\BibitemShut {NoStop}%
\bibitem [{\citenamefont {Fubini}\ and\ \citenamefont {Veneziano}(1969)}]{Fubini:1969qb}%
  \BibitemOpen
  \bibfield  {author} {\bibinfo {author} {\bibfnamefont {S.}~\bibnamefont {Fubini}}\ and\ \bibinfo {author} {\bibfnamefont {G.}~\bibnamefont {Veneziano}},\ }\bibfield  {title} {\bibinfo {title} {{Level structure of dual-resonance models}},\ }\href {https://doi.org/10.1007/BF02758835} {\bibfield  {journal} {\bibinfo  {journal} {Nuovo Cim. A}\ }\textbf {\bibinfo {volume} {64}},\ \bibinfo {pages} {811} (\bibinfo {year} {1969})}\BibitemShut {NoStop}%
\bibitem [{\citenamefont {Huang}\ and\ \citenamefont {Weinberg}(1970)}]{Huang:1970iq}%
  \BibitemOpen
  \bibfield  {author} {\bibinfo {author} {\bibfnamefont {K.}~\bibnamefont {Huang}}\ and\ \bibinfo {author} {\bibfnamefont {S.}~\bibnamefont {Weinberg}},\ }\bibfield  {title} {\bibinfo {title} {{Ultimate temperature and the early universe}},\ }\href {https://doi.org/10.1103/PhysRevLett.25.895} {\bibfield  {journal} {\bibinfo  {journal} {Phys. Rev. Lett.}\ }\textbf {\bibinfo {volume} {25}},\ \bibinfo {pages} {895} (\bibinfo {year} {1970})}\BibitemShut {NoStop}%
\bibitem [{\citenamefont {Dienes}\ and\ \citenamefont {Cudell}(1994)}]{Dienes:1993mp}%
  \BibitemOpen
  \bibfield  {author} {\bibinfo {author} {\bibfnamefont {K.~R.}\ \bibnamefont {Dienes}}\ and\ \bibinfo {author} {\bibfnamefont {J.-R.}\ \bibnamefont {Cudell}},\ }\bibfield  {title} {\bibinfo {title} {{Conformal invariance and degrees of freedom in the QCD string}},\ }\href {https://doi.org/10.1103/PhysRevLett.72.187} {\bibfield  {journal} {\bibinfo  {journal} {Phys. Rev. Lett.}\ }\textbf {\bibinfo {volume} {72}},\ \bibinfo {pages} {187} (\bibinfo {year} {1994})},\ \Eprint {https://arxiv.org/abs/hep-th/9309126} {arXiv:hep-th/9309126} \BibitemShut {NoStop}%
\bibitem [{\citenamefont {Hagedorn}(1965)}]{Hagedorn:1965st}%
  \BibitemOpen
  \bibfield  {author} {\bibinfo {author} {\bibfnamefont {R.}~\bibnamefont {Hagedorn}},\ }\bibfield  {title} {\bibinfo {title} {{Statistical thermodynamics of strong interactions at high-energies}},\ }\href@noop {} {\bibfield  {journal} {\bibinfo  {journal} {Nuovo Cim. Suppl.}\ }\textbf {\bibinfo {volume} {3}},\ \bibinfo {pages} {147} (\bibinfo {year} {1965})}\BibitemShut {NoStop}%
\bibitem [{\citenamefont {Broniowski}\ and\ \citenamefont {Florkowski}(2000)}]{Broniowski:2000bj}%
  \BibitemOpen
  \bibfield  {author} {\bibinfo {author} {\bibfnamefont {W.}~\bibnamefont {Broniowski}}\ and\ \bibinfo {author} {\bibfnamefont {W.}~\bibnamefont {Florkowski}},\ }\bibfield  {title} {\bibinfo {title} {{Different Hagedorn temperatures for mesons and baryons from experimental mass spectra, compound hadrons, and combinatorial saturation}},\ }\href {https://doi.org/10.1016/S0370-2693(00)00992-8} {\bibfield  {journal} {\bibinfo  {journal} {Phys. Lett. B}\ }\textbf {\bibinfo {volume} {490}},\ \bibinfo {pages} {223} (\bibinfo {year} {2000})},\ \Eprint {https://arxiv.org/abs/hep-ph/0004104} {arXiv:hep-ph/0004104} \BibitemShut {NoStop}%
\bibitem [{\citenamefont {Broniowski}\ \emph {et~al.}(2004)\citenamefont {Broniowski}, \citenamefont {Florkowski},\ and\ \citenamefont {Glozman}}]{Broniowski:2004yh}%
  \BibitemOpen
  \bibfield  {author} {\bibinfo {author} {\bibfnamefont {W.}~\bibnamefont {Broniowski}}, \bibinfo {author} {\bibfnamefont {W.}~\bibnamefont {Florkowski}},\ and\ \bibinfo {author} {\bibfnamefont {L.~Y.}\ \bibnamefont {Glozman}},\ }\bibfield  {title} {\bibinfo {title} {{Update of the Hagedorn mass spectrum}},\ }\href {https://doi.org/10.1103/PhysRevD.70.117503} {\bibfield  {journal} {\bibinfo  {journal} {Phys. Rev. D}\ }\textbf {\bibinfo {volume} {70}},\ \bibinfo {pages} {117503} (\bibinfo {year} {2004})},\ \Eprint {https://arxiv.org/abs/hep-ph/0407290} {arXiv:hep-ph/0407290} \BibitemShut {NoStop}%
\bibitem [{\citenamefont {Noronha-Hostler}\ \emph {et~al.}(2010)\citenamefont {Noronha-Hostler}, \citenamefont {Beitel}, \citenamefont {Greiner},\ and\ \citenamefont {Shovkovy}}]{Noronha-Hostler:2009wof}%
  \BibitemOpen
  \bibfield  {author} {\bibinfo {author} {\bibfnamefont {J.}~\bibnamefont {Noronha-Hostler}}, \bibinfo {author} {\bibfnamefont {M.}~\bibnamefont {Beitel}}, \bibinfo {author} {\bibfnamefont {C.}~\bibnamefont {Greiner}},\ and\ \bibinfo {author} {\bibfnamefont {I.}~\bibnamefont {Shovkovy}},\ }\bibfield  {title} {\bibinfo {title} {{Dynamics of Chemical Equilibrium of Hadronic Matter Close to T(c)}},\ }\href {https://doi.org/10.1103/PhysRevC.81.054909} {\bibfield  {journal} {\bibinfo  {journal} {Phys. Rev. C}\ }\textbf {\bibinfo {volume} {81}},\ \bibinfo {pages} {054909} (\bibinfo {year} {2010})},\ \Eprint {https://arxiv.org/abs/0909.2908} {arXiv:0909.2908 [nucl-th]} \BibitemShut {NoStop}%
\bibitem [{\citenamefont {Majumder}\ and\ \citenamefont {Muller}(2010)}]{Majumder:2010ik}%
  \BibitemOpen
  \bibfield  {author} {\bibinfo {author} {\bibfnamefont {A.}~\bibnamefont {Majumder}}\ and\ \bibinfo {author} {\bibfnamefont {B.}~\bibnamefont {Muller}},\ }\bibfield  {title} {\bibinfo {title} {{Hadron Mass Spectrum from Lattice QCD}},\ }\href {https://doi.org/10.1103/PhysRevLett.105.252002} {\bibfield  {journal} {\bibinfo  {journal} {Phys. Rev. Lett.}\ }\textbf {\bibinfo {volume} {105}},\ \bibinfo {pages} {252002} (\bibinfo {year} {2010})},\ \Eprint {https://arxiv.org/abs/1008.1747} {arXiv:1008.1747 [hep-ph]} \BibitemShut {NoStop}%
\bibitem [{\citenamefont {Cleymans}\ and\ \citenamefont {Worku}(2011)}]{Cleymans:2011fx}%
  \BibitemOpen
  \bibfield  {author} {\bibinfo {author} {\bibfnamefont {J.}~\bibnamefont {Cleymans}}\ and\ \bibinfo {author} {\bibfnamefont {D.}~\bibnamefont {Worku}},\ }\bibfield  {title} {\bibinfo {title} {{The Hagedorn temperature Revisited}},\ }\href {https://doi.org/10.1142/S0217732311035584} {\bibfield  {journal} {\bibinfo  {journal} {Mod. Phys. Lett. A}\ }\textbf {\bibinfo {volume} {26}},\ \bibinfo {pages} {1197} (\bibinfo {year} {2011})},\ \Eprint {https://arxiv.org/abs/1103.1463} {arXiv:1103.1463 [hep-ph]} \BibitemShut {NoStop}%
\bibitem [{\citenamefont {Lo}\ \emph {et~al.}(2015)\citenamefont {Lo}, \citenamefont {Marczenko}, \citenamefont {Redlich},\ and\ \citenamefont {Sasaki}}]{Lo:2015cca}%
  \BibitemOpen
  \bibfield  {author} {\bibinfo {author} {\bibfnamefont {P.~M.}\ \bibnamefont {Lo}}, \bibinfo {author} {\bibfnamefont {M.}~\bibnamefont {Marczenko}}, \bibinfo {author} {\bibfnamefont {K.}~\bibnamefont {Redlich}},\ and\ \bibinfo {author} {\bibfnamefont {C.}~\bibnamefont {Sasaki}},\ }\bibfield  {title} {\bibinfo {title} {{Matching the Hagedorn mass spectrum with Lattice QCD results}},\ }\href {https://doi.org/10.1103/PhysRevC.92.055206} {\bibfield  {journal} {\bibinfo  {journal} {Phys. Rev. C}\ }\textbf {\bibinfo {volume} {92}},\ \bibinfo {pages} {055206} (\bibinfo {year} {2015})},\ \Eprint {https://arxiv.org/abs/1507.06398} {arXiv:1507.06398 [nucl-th]} \BibitemShut {NoStop}%
\bibitem [{\citenamefont {Man~Lo}\ \emph {et~al.}(2016)\citenamefont {Man~Lo}, \citenamefont {Marczenko}, \citenamefont {Redlich},\ and\ \citenamefont {Sasaki}}]{ManLo:2016pgd}%
  \BibitemOpen
  \bibfield  {author} {\bibinfo {author} {\bibfnamefont {P.}~\bibnamefont {Man~Lo}}, \bibinfo {author} {\bibfnamefont {M.}~\bibnamefont {Marczenko}}, \bibinfo {author} {\bibfnamefont {K.}~\bibnamefont {Redlich}},\ and\ \bibinfo {author} {\bibfnamefont {C.}~\bibnamefont {Sasaki}},\ }\bibfield  {title} {\bibinfo {title} {{Missing baryonic resonances in the Hagedorn spectrum}},\ }\href {https://doi.org/10.1140/epja/i2016-16235-6} {\bibfield  {journal} {\bibinfo  {journal} {Eur. Phys. J. A}\ }\textbf {\bibinfo {volume} {52}},\ \bibinfo {pages} {235} (\bibinfo {year} {2016})}\BibitemShut {NoStop}%
\bibitem [{\citenamefont {Meyer}(2009)}]{Meyer:2009tq}%
  \BibitemOpen
  \bibfield  {author} {\bibinfo {author} {\bibfnamefont {H.~B.}\ \bibnamefont {Meyer}},\ }\bibfield  {title} {\bibinfo {title} {{High-Precision Thermodynamics and Hagedorn Density of States}},\ }\href {https://doi.org/10.1103/PhysRevD.80.051502} {\bibfield  {journal} {\bibinfo  {journal} {Phys. Rev. D}\ }\textbf {\bibinfo {volume} {80}},\ \bibinfo {pages} {051502} (\bibinfo {year} {2009})},\ \Eprint {https://arxiv.org/abs/0905.4229} {arXiv:0905.4229 [hep-lat]} \BibitemShut {NoStop}%
\bibitem [{\citenamefont {Nambu}(1974)}]{Nambu:1974zg}%
  \BibitemOpen
  \bibfield  {author} {\bibinfo {author} {\bibfnamefont {Y.}~\bibnamefont {Nambu}},\ }\bibfield  {title} {\bibinfo {title} {{Strings, Monopoles and Gauge Fields}},\ }\href {https://doi.org/10.1103/PhysRevD.10.4262} {\bibfield  {journal} {\bibinfo  {journal} {Phys. Rev. D}\ }\textbf {\bibinfo {volume} {10}},\ \bibinfo {pages} {4262} (\bibinfo {year} {1974})}\BibitemShut {NoStop}%
\bibitem [{\citenamefont {Bars}\ and\ \citenamefont {Hanson}(1976)}]{Bars:1975dd}%
  \BibitemOpen
  \bibfield  {author} {\bibinfo {author} {\bibfnamefont {I.}~\bibnamefont {Bars}}\ and\ \bibinfo {author} {\bibfnamefont {A.~J.}\ \bibnamefont {Hanson}},\ }\bibfield  {title} {\bibinfo {title} {{Quarks at the Ends of the String}},\ }\href {https://doi.org/10.1103/PhysRevD.13.1744} {\bibfield  {journal} {\bibinfo  {journal} {Phys. Rev. D}\ }\textbf {\bibinfo {volume} {13}},\ \bibinfo {pages} {1744} (\bibinfo {year} {1976})}\BibitemShut {NoStop}%
\bibitem [{\citenamefont {Selem}\ and\ \citenamefont {Wilczek}(2006)}]{Selem:2006nd}%
  \BibitemOpen
  \bibfield  {author} {\bibinfo {author} {\bibfnamefont {A.}~\bibnamefont {Selem}}\ and\ \bibinfo {author} {\bibfnamefont {F.}~\bibnamefont {Wilczek}},\ }\bibfield  {title} {\bibinfo {title} {{Hadron systematics and emergent diquarks}},\ }in\ \href {https://doi.org/10.1142/9789812773524_0030} {\emph {\bibinfo {booktitle} {{Ringberg Workshop on New Trends in HERA Physics 2005}}}}\ (\bibinfo {year} {2006})\ pp.\ \bibinfo {pages} {337--356},\ \Eprint {https://arxiv.org/abs/hep-ph/0602128} {arXiv:hep-ph/0602128} \BibitemShut {NoStop}%
\bibitem [{\citenamefont {Lucini}\ \emph {et~al.}(2004)\citenamefont {Lucini}, \citenamefont {Teper},\ and\ \citenamefont {Wenger}}]{Lucini:2003zr}%
  \BibitemOpen
  \bibfield  {author} {\bibinfo {author} {\bibfnamefont {B.}~\bibnamefont {Lucini}}, \bibinfo {author} {\bibfnamefont {M.}~\bibnamefont {Teper}},\ and\ \bibinfo {author} {\bibfnamefont {U.}~\bibnamefont {Wenger}},\ }\bibfield  {title} {\bibinfo {title} {{The High temperature phase transition in SU(N) gauge theories}},\ }\href {https://doi.org/10.1088/1126-6708/2004/01/061} {\bibfield  {journal} {\bibinfo  {journal} {JHEP}\ }\textbf {\bibinfo {volume} {01}},\ \bibinfo {pages} {061}},\ \Eprint {https://arxiv.org/abs/hep-lat/0307017} {arXiv:hep-lat/0307017} \BibitemShut {NoStop}%
\bibitem [{\citenamefont {Athenodorou}\ and\ \citenamefont {Teper}(2020)}]{Athenodorou:2020ani}%
  \BibitemOpen
  \bibfield  {author} {\bibinfo {author} {\bibfnamefont {A.}~\bibnamefont {Athenodorou}}\ and\ \bibinfo {author} {\bibfnamefont {M.}~\bibnamefont {Teper}},\ }\bibfield  {title} {\bibinfo {title} {{The glueball spectrum of SU(3) gauge theory in 3 + 1 dimensions}},\ }\href {https://doi.org/10.1007/JHEP11(2020)172} {\bibfield  {journal} {\bibinfo  {journal} {JHEP}\ }\textbf {\bibinfo {volume} {11}},\ \bibinfo {pages} {172}},\ \Eprint {https://arxiv.org/abs/2007.06422} {arXiv:2007.06422 [hep-lat]} \BibitemShut {NoStop}%
\bibitem [{\citenamefont {Brambilla}\ \emph {et~al.}(2023)\citenamefont {Brambilla}, \citenamefont {Delgado}, \citenamefont {Kronfeld}, \citenamefont {Leino}, \citenamefont {Petreczky}, \citenamefont {Steinbei{\ss}er}, \citenamefont {Vairo},\ and\ \citenamefont {Weber}}]{Brambilla:2022het}%
  \BibitemOpen
  \bibfield  {author} {\bibinfo {author} {\bibfnamefont {N.}~\bibnamefont {Brambilla}}, \bibinfo {author} {\bibfnamefont {R.~L.}\ \bibnamefont {Delgado}}, \bibinfo {author} {\bibfnamefont {A.~S.}\ \bibnamefont {Kronfeld}}, \bibinfo {author} {\bibfnamefont {V.}~\bibnamefont {Leino}}, \bibinfo {author} {\bibfnamefont {P.}~\bibnamefont {Petreczky}}, \bibinfo {author} {\bibfnamefont {S.}~\bibnamefont {Steinbei{\ss}er}}, \bibinfo {author} {\bibfnamefont {A.}~\bibnamefont {Vairo}},\ and\ \bibinfo {author} {\bibfnamefont {J.~H.}\ \bibnamefont {Weber}} (\bibinfo {collaboration} {TUMQCD}),\ }\bibfield  {title} {\bibinfo {title} {{Static energy in ($2+1+1$)-flavor lattice QCD: Scale setting and charm effects}},\ }\href {https://doi.org/10.1103/PhysRevD.107.074503} {\bibfield  {journal} {\bibinfo  {journal} {Phys. Rev. D}\ }\textbf {\bibinfo {volume} {107}},\ \bibinfo {pages} {074503} (\bibinfo {year} {2023})},\ \Eprint {https://arxiv.org/abs/2206.03156} {arXiv:2206.03156 [hep-lat]} \BibitemShut {NoStop}%
\bibitem [{\citenamefont {Olsson}\ and\ \citenamefont {Veseli}(1995)}]{Olsson:1994dg}%
  \BibitemOpen
  \bibfield  {author} {\bibinfo {author} {\bibfnamefont {M.~G.}\ \bibnamefont {Olsson}}\ and\ \bibinfo {author} {\bibfnamefont {S.}~\bibnamefont {Veseli}},\ }\bibfield  {title} {\bibinfo {title} {{The Asymmetric flux tube}},\ }\href {https://doi.org/10.1103/PhysRevD.51.3578} {\bibfield  {journal} {\bibinfo  {journal} {Phys. Rev. D}\ }\textbf {\bibinfo {volume} {51}},\ \bibinfo {pages} {3578} (\bibinfo {year} {1995})}\BibitemShut {NoStop}%
\bibitem [{\citenamefont {Filipponi}\ and\ \citenamefont {Srivastava}(1998)}]{Filipponi:1997vf}%
  \BibitemOpen
  \bibfield  {author} {\bibinfo {author} {\bibfnamefont {S.}~\bibnamefont {Filipponi}}\ and\ \bibinfo {author} {\bibfnamefont {Y.}~\bibnamefont {Srivastava}},\ }\bibfield  {title} {\bibinfo {title} {{Hadronic masses and Regge trajectories}},\ }\href {https://doi.org/10.1103/PhysRevD.58.016003} {\bibfield  {journal} {\bibinfo  {journal} {Phys. Rev. D}\ }\textbf {\bibinfo {volume} {58}},\ \bibinfo {pages} {016003} (\bibinfo {year} {1998})},\ \Eprint {https://arxiv.org/abs/hep-ph/9712204} {arXiv:hep-ph/9712204} \BibitemShut {NoStop}%
\bibitem [{\citenamefont {Sonnenschein}\ and\ \citenamefont {Weissman}(2014)}]{Sonnenschein:2014jwa}%
  \BibitemOpen
  \bibfield  {author} {\bibinfo {author} {\bibfnamefont {J.}~\bibnamefont {Sonnenschein}}\ and\ \bibinfo {author} {\bibfnamefont {D.}~\bibnamefont {Weissman}},\ }\bibfield  {title} {\bibinfo {title} {{Rotating strings confronting PDG mesons}},\ }\href {https://doi.org/10.1007/JHEP08(2014)013} {\bibfield  {journal} {\bibinfo  {journal} {JHEP}\ }\textbf {\bibinfo {volume} {08}},\ \bibinfo {pages} {013}},\ \Eprint {https://arxiv.org/abs/1402.5603} {arXiv:1402.5603 [hep-ph]} \BibitemShut {NoStop}%
\bibitem [{\citenamefont {Chen}\ \emph {et~al.}(2018)\citenamefont {Chen}, \citenamefont {Dong}, \citenamefont {Liu}, \citenamefont {L{\"u}},\ and\ \citenamefont {Matsuki}}]{Chen:2017fcs}%
  \BibitemOpen
  \bibfield  {author} {\bibinfo {author} {\bibfnamefont {K.}~\bibnamefont {Chen}}, \bibinfo {author} {\bibfnamefont {Y.}~\bibnamefont {Dong}}, \bibinfo {author} {\bibfnamefont {X.}~\bibnamefont {Liu}}, \bibinfo {author} {\bibfnamefont {Q.-F.}\ \bibnamefont {L{\"u}}},\ and\ \bibinfo {author} {\bibfnamefont {T.}~\bibnamefont {Matsuki}},\ }\bibfield  {title} {\bibinfo {title} {{Regge-like relation and a universal description of heavy{\textendash}light systems}},\ }\href {https://doi.org/10.1140/epjc/s10052-017-5512-3} {\bibfield  {journal} {\bibinfo  {journal} {Eur. Phys. J. C}\ }\textbf {\bibinfo {volume} {78}},\ \bibinfo {pages} {20} (\bibinfo {year} {2018})},\ \Eprint {https://arxiv.org/abs/1709.07196} {arXiv:1709.07196 [hep-ph]} \BibitemShut {NoStop}%
\bibitem [{\citenamefont {Artru}(1975)}]{Artru:1974zn}%
  \BibitemOpen
  \bibfield  {author} {\bibinfo {author} {\bibfnamefont {X.}~\bibnamefont {Artru}},\ }\bibfield  {title} {\bibinfo {title} {{String Model with Baryons: Topology, Classical Motion}},\ }\href {https://doi.org/10.1016/0550-3213(75)90019-X} {\bibfield  {journal} {\bibinfo  {journal} {Nucl. Phys. B}\ }\textbf {\bibinfo {volume} {85}},\ \bibinfo {pages} {442} (\bibinfo {year} {1975})}\BibitemShut {NoStop}%
\bibitem [{\citenamefont {Collins}\ \emph {et~al.}(1975)\citenamefont {Collins}, \citenamefont {Hopkinson},\ and\ \citenamefont {Tucker}}]{Collins:1975yt}%
  \BibitemOpen
  \bibfield  {author} {\bibinfo {author} {\bibfnamefont {P.~A.}\ \bibnamefont {Collins}}, \bibinfo {author} {\bibfnamefont {J.~F.~L.}\ \bibnamefont {Hopkinson}},\ and\ \bibinfo {author} {\bibfnamefont {R.~W.}\ \bibnamefont {Tucker}},\ }\bibfield  {title} {\bibinfo {title} {{Classical Solutions for the Relativistic Three String Baryon Problem}},\ }\href {https://doi.org/10.1016/0550-3213(75)90230-8} {\bibfield  {journal} {\bibinfo  {journal} {Nucl. Phys. B}\ }\textbf {\bibinfo {volume} {100}},\ \bibinfo {pages} {157} (\bibinfo {year} {1975})}\BibitemShut {NoStop}%
\bibitem [{\citenamefont {Kiriyama}(1976)}]{Kiriyama:1976ue}%
  \BibitemOpen
  \bibfield  {author} {\bibinfo {author} {\bibfnamefont {K.}~\bibnamefont {Kiriyama}},\ }\bibfield  {title} {\bibinfo {title} {{A String Model with Several Arms}},\ }\href {https://doi.org/10.1007/BF02721008} {\bibfield  {journal} {\bibinfo  {journal} {Lett. Nuovo Cim.}\ }\textbf {\bibinfo {volume} {16}},\ \bibinfo {pages} {467} (\bibinfo {year} {1976})}\BibitemShut {NoStop}%
\bibitem [{\citenamefont {Sundermeyer}\ and\ \citenamefont {de~la Torre}(1977)}]{Sundermeyer:1976df}%
  \BibitemOpen
  \bibfield  {author} {\bibinfo {author} {\bibfnamefont {K.}~\bibnamefont {Sundermeyer}}\ and\ \bibinfo {author} {\bibfnamefont {A.}~\bibnamefont {de~la Torre}},\ }\bibfield  {title} {\bibinfo {title} {{Quantization of a Three-String}},\ }\href {https://doi.org/10.1103/PhysRevD.15.1745} {\bibfield  {journal} {\bibinfo  {journal} {Phys. Rev. D}\ }\textbf {\bibinfo {volume} {15}},\ \bibinfo {pages} {1745} (\bibinfo {year} {1977})}\BibitemShut {NoStop}%
\bibitem [{\citenamefont {Rossi}\ and\ \citenamefont {Veneziano}(1977)}]{Rossi:1977cy}%
  \BibitemOpen
  \bibfield  {author} {\bibinfo {author} {\bibfnamefont {G.~C.}\ \bibnamefont {Rossi}}\ and\ \bibinfo {author} {\bibfnamefont {G.}~\bibnamefont {Veneziano}},\ }\bibfield  {title} {\bibinfo {title} {{A Possible Description of Baryon Dynamics in Dual and Gauge Theories}},\ }\href {https://doi.org/10.1016/0550-3213(77)90178-X} {\bibfield  {journal} {\bibinfo  {journal} {Nucl. Phys. B}\ }\textbf {\bibinfo {volume} {123}},\ \bibinfo {pages} {507} (\bibinfo {year} {1977})}\BibitemShut {NoStop}%
\bibitem [{\citenamefont {Asano}\ and\ \citenamefont {Kato}(2022)}]{Asano:2022wou}%
  \BibitemOpen
  \bibfield  {author} {\bibinfo {author} {\bibfnamefont {M.}~\bibnamefont {Asano}}\ and\ \bibinfo {author} {\bibfnamefont {M.}~\bibnamefont {Kato}},\ }\bibfield  {title} {\bibinfo {title} {{Quantizing a multi-pronged open string junction}},\ }\href {https://doi.org/10.1093/ptep/ptac125} {\bibfield  {journal} {\bibinfo  {journal} {PTEP}\ }\textbf {\bibinfo {volume} {2022}},\ \bibinfo {pages} {103B01} (\bibinfo {year} {2022})},\ \Eprint {https://arxiv.org/abs/2207.00131} {arXiv:2207.00131 [hep-th]} \BibitemShut {NoStop}%
\bibitem [{\citenamefont {Frenklakh}\ \emph {et~al.}(2024)\citenamefont {Frenklakh}, \citenamefont {Kharzeev},\ and\ \citenamefont {Li}}]{Frenklakh:2023pwy}%
  \BibitemOpen
  \bibfield  {author} {\bibinfo {author} {\bibfnamefont {D.}~\bibnamefont {Frenklakh}}, \bibinfo {author} {\bibfnamefont {D.~E.}\ \bibnamefont {Kharzeev}},\ and\ \bibinfo {author} {\bibfnamefont {W.}~\bibnamefont {Li}},\ }\bibfield  {title} {\bibinfo {title} {{Signatures of baryon junctions in semi-inclusive deep inelastic scattering}},\ }\href {https://doi.org/10.1016/j.physletb.2024.138680} {\bibfield  {journal} {\bibinfo  {journal} {Phys. Lett. B}\ }\textbf {\bibinfo {volume} {853}},\ \bibinfo {pages} {138680} (\bibinfo {year} {2024})},\ \Eprint {https://arxiv.org/abs/2312.15039} {arXiv:2312.15039 [hep-ph]} \BibitemShut {NoStop}%
\bibitem [{\citenamefont {Altmann}\ and\ \citenamefont {Skands}(2024)}]{Altmann:2024odn}%
  \BibitemOpen
  \bibfield  {author} {\bibinfo {author} {\bibfnamefont {J.}~\bibnamefont {Altmann}}\ and\ \bibinfo {author} {\bibfnamefont {P.}~\bibnamefont {Skands}},\ }\bibfield  {title} {\bibinfo {title} {{String junctions revisited}},\ }\href {https://doi.org/10.1007/JHEP07(2024)238} {\bibfield  {journal} {\bibinfo  {journal} {JHEP}\ }\textbf {\bibinfo {volume} {07}},\ \bibinfo {pages} {238}},\ \Eprint {https://arxiv.org/abs/2404.12040} {arXiv:2404.12040 [hep-ph]} \BibitemShut {NoStop}%
\bibitem [{\citenamefont {Komargodski}\ and\ \citenamefont {Zhong}(2024)}]{Komargodski:2024swh}%
  \BibitemOpen
  \bibfield  {author} {\bibinfo {author} {\bibfnamefont {Z.}~\bibnamefont {Komargodski}}\ and\ \bibinfo {author} {\bibfnamefont {S.}~\bibnamefont {Zhong}},\ }\bibfield  {title} {\bibinfo {title} {{Baryon junction and string interactions}},\ }\href {https://doi.org/10.1103/PhysRevD.110.056018} {\bibfield  {journal} {\bibinfo  {journal} {Phys. Rev. D}\ }\textbf {\bibinfo {volume} {110}},\ \bibinfo {pages} {056018} (\bibinfo {year} {2024})},\ \Eprint {https://arxiv.org/abs/2405.12005} {arXiv:2405.12005 [hep-th]} \BibitemShut {NoStop}%
\bibitem [{\citenamefont {Takahashi}\ \emph {et~al.}(2001)\citenamefont {Takahashi}, \citenamefont {Matsufuru}, \citenamefont {Nemoto},\ and\ \citenamefont {Suganuma}}]{Takahashi:2000te}%
  \BibitemOpen
  \bibfield  {author} {\bibinfo {author} {\bibfnamefont {T.~T.}\ \bibnamefont {Takahashi}}, \bibinfo {author} {\bibfnamefont {H.}~\bibnamefont {Matsufuru}}, \bibinfo {author} {\bibfnamefont {Y.}~\bibnamefont {Nemoto}},\ and\ \bibinfo {author} {\bibfnamefont {H.}~\bibnamefont {Suganuma}},\ }\bibfield  {title} {\bibinfo {title} {{The Three quark potential in the SU(3) lattice QCD}},\ }\href {https://doi.org/10.1103/PhysRevLett.86.18} {\bibfield  {journal} {\bibinfo  {journal} {Phys. Rev. Lett.}\ }\textbf {\bibinfo {volume} {86}},\ \bibinfo {pages} {18} (\bibinfo {year} {2001})},\ \Eprint {https://arxiv.org/abs/hep-lat/0006005} {arXiv:hep-lat/0006005} \BibitemShut {NoStop}%
\bibitem [{\citenamefont {Takahashi}\ \emph {et~al.}(2002)\citenamefont {Takahashi}, \citenamefont {Suganuma}, \citenamefont {Nemoto},\ and\ \citenamefont {Matsufuru}}]{Takahashi:2002bw}%
  \BibitemOpen
  \bibfield  {author} {\bibinfo {author} {\bibfnamefont {T.~T.}\ \bibnamefont {Takahashi}}, \bibinfo {author} {\bibfnamefont {H.}~\bibnamefont {Suganuma}}, \bibinfo {author} {\bibfnamefont {Y.}~\bibnamefont {Nemoto}},\ and\ \bibinfo {author} {\bibfnamefont {H.}~\bibnamefont {Matsufuru}},\ }\bibfield  {title} {\bibinfo {title} {{Detailed analysis of the three quark potential in SU(3) lattice QCD}},\ }\href {https://doi.org/10.1103/PhysRevD.65.114509} {\bibfield  {journal} {\bibinfo  {journal} {Phys. Rev. D}\ }\textbf {\bibinfo {volume} {65}},\ \bibinfo {pages} {114509} (\bibinfo {year} {2002})},\ \Eprint {https://arxiv.org/abs/hep-lat/0204011} {arXiv:hep-lat/0204011} \BibitemShut {NoStop}%
\bibitem [{\citenamefont {Alexandrou}\ \emph {et~al.}(2003)\citenamefont {Alexandrou}, \citenamefont {de~Forcrand},\ and\ \citenamefont {Jahn}}]{Alexandrou:2002sn}%
  \BibitemOpen
  \bibfield  {author} {\bibinfo {author} {\bibfnamefont {C.}~\bibnamefont {Alexandrou}}, \bibinfo {author} {\bibfnamefont {P.}~\bibnamefont {de~Forcrand}},\ and\ \bibinfo {author} {\bibfnamefont {O.}~\bibnamefont {Jahn}},\ }\bibfield  {title} {\bibinfo {title} {{The Ground state of three quarks}},\ }\href {https://doi.org/10.1016/S0920-5632(03)01659-1} {\bibfield  {journal} {\bibinfo  {journal} {Nucl. Phys. B Proc. Suppl.}\ }\textbf {\bibinfo {volume} {119}},\ \bibinfo {pages} {667} (\bibinfo {year} {2003})},\ \Eprint {https://arxiv.org/abs/hep-lat/0209062} {arXiv:hep-lat/0209062} \BibitemShut {NoStop}%
\bibitem [{\citenamefont {Bissey}\ \emph {et~al.}(2007)\citenamefont {Bissey}, \citenamefont {Cao}, \citenamefont {Kitson}, \citenamefont {Signal}, \citenamefont {Leinweber}, \citenamefont {Lasscock},\ and\ \citenamefont {Williams}}]{Bissey:2006bz}%
  \BibitemOpen
  \bibfield  {author} {\bibinfo {author} {\bibfnamefont {F.}~\bibnamefont {Bissey}}, \bibinfo {author} {\bibfnamefont {F.-G.}\ \bibnamefont {Cao}}, \bibinfo {author} {\bibfnamefont {A.~R.}\ \bibnamefont {Kitson}}, \bibinfo {author} {\bibfnamefont {A.~I.}\ \bibnamefont {Signal}}, \bibinfo {author} {\bibfnamefont {D.~B.}\ \bibnamefont {Leinweber}}, \bibinfo {author} {\bibfnamefont {B.~G.}\ \bibnamefont {Lasscock}},\ and\ \bibinfo {author} {\bibfnamefont {A.~G.}\ \bibnamefont {Williams}},\ }\bibfield  {title} {\bibinfo {title} {{Gluon flux-tube distribution and linear confinement in baryons}},\ }\href {https://doi.org/10.1103/PhysRevD.76.114512} {\bibfield  {journal} {\bibinfo  {journal} {Phys. Rev. D}\ }\textbf {\bibinfo {volume} {76}},\ \bibinfo {pages} {114512} (\bibinfo {year} {2007})},\ \Eprint {https://arxiv.org/abs/hep-lat/0606016} {arXiv:hep-lat/0606016} \BibitemShut {NoStop}%
\bibitem [{\citenamefont {Kharzeev}(1996)}]{Kharzeev:1996sq}%
  \BibitemOpen
  \bibfield  {author} {\bibinfo {author} {\bibfnamefont {D.}~\bibnamefont {Kharzeev}},\ }\bibfield  {title} {\bibinfo {title} {{Can gluons trace baryon number?}},\ }\href {https://doi.org/10.1016/0370-2693(96)00435-2} {\bibfield  {journal} {\bibinfo  {journal} {Phys. Lett. B}\ }\textbf {\bibinfo {volume} {378}},\ \bibinfo {pages} {238} (\bibinfo {year} {1996})},\ \Eprint {https://arxiv.org/abs/nucl-th/9602027} {arXiv:nucl-th/9602027} \BibitemShut {NoStop}%
\bibitem [{\citenamefont {Aboona}\ \emph {et~al.}(2026)\citenamefont {Aboona} \emph {et~al.}}]{STAR:2024lvy}%
  \BibitemOpen
  \bibfield  {author} {\bibinfo {author} {\bibfnamefont {B.~E.}\ \bibnamefont {Aboona}} \emph {et~al.} (\bibinfo {collaboration} {STAR}),\ }\bibfield  {title} {\bibinfo {title} {{Tracking the baryon number with nuclear collisions}},\ }\href {https://doi.org/10.1126/science.ads5962} {\bibfield  {journal} {\bibinfo  {journal} {Science}\ }\textbf {\bibinfo {volume} {393}},\ \bibinfo {pages} {ads5962} (\bibinfo {year} {2026})},\ \Eprint {https://arxiv.org/abs/2408.15441} {arXiv:2408.15441 [nucl-ex]} \BibitemShut {NoStop}%
\bibitem [{\citenamefont {Green}\ \emph {et~al.}(1988)\citenamefont {Green}, \citenamefont {Schwarz},\ and\ \citenamefont {Witten}}]{Green:1987sp}%
  \BibitemOpen
  \bibfield  {author} {\bibinfo {author} {\bibfnamefont {M.~B.}\ \bibnamefont {Green}}, \bibinfo {author} {\bibfnamefont {J.~H.}\ \bibnamefont {Schwarz}},\ and\ \bibinfo {author} {\bibfnamefont {E.}~\bibnamefont {Witten}},\ }\href@noop {} {\emph {\bibinfo {title} {{SUPERSTRING THEORY. VOL. 1: INTRODUCTION}}}},\ Cambridge Monographs on Mathematical Physics\ (\bibinfo {year} {1988})\BibitemShut {NoStop}%
\bibitem [{\citenamefont {Tong}(2009)}]{Tong:2009np}%
  \BibitemOpen
  \bibfield  {author} {\bibinfo {author} {\bibfnamefont {D.}~\bibnamefont {Tong}},\ }\bibfield  {title} {\bibinfo {title} {{String Theory}},\ }\href@noop {} {\  (\bibinfo {year} {2009})},\ \Eprint {https://arxiv.org/abs/0908.0333} {arXiv:0908.0333 [hep-th]} \BibitemShut {NoStop}%
\bibitem [{\citenamefont {Gursoy}\ and\ \citenamefont {Saclioglu}(2001)}]{Gursoy:2001bb}%
  \BibitemOpen
  \bibfield  {author} {\bibinfo {author} {\bibfnamefont {U.}~\bibnamefont {Gursoy}}\ and\ \bibinfo {author} {\bibfnamefont {C.}~\bibnamefont {Saclioglu}},\ }\bibfield  {title} {\bibinfo {title} {{An open bosonic string with one end fixed}},\ }\href {https://doi.org/10.1088/0264-9381/18/14/305} {\bibfield  {journal} {\bibinfo  {journal} {Class. Quant. Grav.}\ }\textbf {\bibinfo {volume} {18}},\ \bibinfo {pages} {2665} (\bibinfo {year} {2001})},\ \Eprint {https://arxiv.org/abs/hep-th/0102003} {arXiv:hep-th/0102003} \BibitemShut {NoStop}%
\bibitem [{\citenamefont {Navas}\ \emph {et~al.}(2024)\citenamefont {Navas} \emph {et~al.}}]{ParticleDataGroup:2024cfk}%
  \BibitemOpen
  \bibfield  {author} {\bibinfo {author} {\bibfnamefont {S.}~\bibnamefont {Navas}} \emph {et~al.} (\bibinfo {collaboration} {Particle Data Group}),\ }\bibfield  {title} {\bibinfo {title} {{Review of particle physics}},\ }\href {https://doi.org/10.1103/PhysRevD.110.030001} {\bibfield  {journal} {\bibinfo  {journal} {Phys. Rev. D}\ }\textbf {\bibinfo {volume} {110}},\ \bibinfo {pages} {030001} (\bibinfo {year} {2024})}\BibitemShut {NoStop}%
\bibitem [{\citenamefont {Cabibbo}\ and\ \citenamefont {Parisi}(1975)}]{Cabibbo:1975ig}%
  \BibitemOpen
  \bibfield  {author} {\bibinfo {author} {\bibfnamefont {N.}~\bibnamefont {Cabibbo}}\ and\ \bibinfo {author} {\bibfnamefont {G.}~\bibnamefont {Parisi}},\ }\bibfield  {title} {\bibinfo {title} {{Exponential Hadronic Spectrum and Quark Liberation}},\ }\href {https://doi.org/10.1016/0370-2693(75)90158-6} {\bibfield  {journal} {\bibinfo  {journal} {Phys. Lett. B}\ }\textbf {\bibinfo {volume} {59}},\ \bibinfo {pages} {67} (\bibinfo {year} {1975})}\BibitemShut {NoStop}%
\bibitem [{\citenamefont {Borsanyi}\ \emph {et~al.}(2022)\citenamefont {Borsanyi}, \citenamefont {R.}, \citenamefont {Fodor}, \citenamefont {Godzieba}, \citenamefont {Parotto},\ and\ \citenamefont {Sexty}}]{Borsanyi:2022xml}%
  \BibitemOpen
  \bibfield  {author} {\bibinfo {author} {\bibfnamefont {S.}~\bibnamefont {Borsanyi}}, \bibinfo {author} {\bibfnamefont {K.}~\bibnamefont {R.}}, \bibinfo {author} {\bibfnamefont {Z.}~\bibnamefont {Fodor}}, \bibinfo {author} {\bibfnamefont {D.~A.}\ \bibnamefont {Godzieba}}, \bibinfo {author} {\bibfnamefont {P.}~\bibnamefont {Parotto}},\ and\ \bibinfo {author} {\bibfnamefont {D.}~\bibnamefont {Sexty}},\ }\bibfield  {title} {\bibinfo {title} {{Precision study of the continuum SU(3) Yang-Mills theory: How to use parallel tempering to improve on supercritical slowing down for first order phase transitions}},\ }\href {https://doi.org/10.1103/PhysRevD.105.074513} {\bibfield  {journal} {\bibinfo  {journal} {Phys. Rev. D}\ }\textbf {\bibinfo {volume} {105}},\ \bibinfo {pages} {074513} (\bibinfo {year} {2022})},\ \Eprint {https://arxiv.org/abs/2202.05234} {arXiv:2202.05234 [hep-lat]} \BibitemShut {NoStop}%
\bibitem [{\citenamefont {Giusti}\ \emph {et~al.}(2025)\citenamefont {Giusti}, \citenamefont {Hirasawa}, \citenamefont {Pepe},\ and\ \citenamefont {Virz{\`\i}}}]{Giusti:2025fxu}%
  \BibitemOpen
  \bibfield  {author} {\bibinfo {author} {\bibfnamefont {L.}~\bibnamefont {Giusti}}, \bibinfo {author} {\bibfnamefont {M.}~\bibnamefont {Hirasawa}}, \bibinfo {author} {\bibfnamefont {M.}~\bibnamefont {Pepe}},\ and\ \bibinfo {author} {\bibfnamefont {L.}~\bibnamefont {Virz{\`\i}}},\ }\bibfield  {title} {\bibinfo {title} {{A precise study of the thermodynamic properties of the SU(3) Yang-Mills theory across the deconfinement transition}},\ }\href {https://doi.org/10.1016/j.physletb.2025.139775} {\bibfield  {journal} {\bibinfo  {journal} {Phys. Lett. B}\ }\textbf {\bibinfo {volume} {868}},\ \bibinfo {pages} {139775} (\bibinfo {year} {2025})},\ \Eprint {https://arxiv.org/abs/2501.10284} {arXiv:2501.10284 [hep-lat]} \BibitemShut {NoStop}%
\bibitem [{\citenamefont {Svetitsky}(1986)}]{Svetitsky:1985ye}%
  \BibitemOpen
  \bibfield  {author} {\bibinfo {author} {\bibfnamefont {B.}~\bibnamefont {Svetitsky}},\ }\bibfield  {title} {\bibinfo {title} {{Symmetry Aspects of Finite Temperature Confinement Transitions}},\ }\href {https://doi.org/10.1016/0370-1573(86)90014-1} {\bibfield  {journal} {\bibinfo  {journal} {Phys. Rept.}\ }\textbf {\bibinfo {volume} {132}},\ \bibinfo {pages} {1} (\bibinfo {year} {1986})}\BibitemShut {NoStop}%
\end{thebibliography}%
\end{document}